\documentclass[10pt, conference, letterpaper]{IEEEtran}
\usepackage{graphicx}
\graphicspath{{./Pictures/}} % Specifies the directory where pictures are stored
\usepackage{booktabs}
\usepackage{amsmath}
\usepackage{url}
 \usepackage{mathrsfs}
\usepackage{amssymb}

\usepackage{multirow}
\usepackage{array}
\newcolumntype{L}[1]{>{\raggedright\let\newline\\\arraybackslash\hspace{0pt}}m{#1}}
\newcolumntype{C}[1]{>{\centering\let\newline\\\arraybackslash\hspace{0pt}}m{#1}}
\newcolumntype{R}[1]{>{\raggedleft\let\newline\\\arraybackslash\hspace{0pt}}m{#1}}

\usepackage{bigstrut}
\usepackage{cite}
\ifCLASSOPTIONcompsoc
\usepackage[caption=false,font=normalsize,labelfont=sf,textfont=sf]{subfig}
\else
\usepackage[caption=false,font=footnotesize]{subfig}
\fi

\usepackage{color,xcolor,colortbl}

\usepackage{footnote}
\makesavenoteenv{table}

\usepackage{algorithm}
\usepackage{algpseudocode}
\algnewcommand\algorithmicinput{\textbf{INPUT:}}
\algnewcommand\INPUT{\item[\algorithmicinput]}
\algnewcommand{\algorithmicoutput}{\textbf{OUTPUT:}}
\algnewcommand\OUTPUT{\item[\algorithmicoutput]}
\algrenewcommand{\algorithmiccomment}[1]{\hskip3em$\%$ #1}

\begin{document}

\title{Radio Frequency Fingerprint Identification for LoRa Using Spectrogram and CNN}
%Radio Frequency Fingerprinting Based On Deep Learning and Short-time Fourier Transform

%\author{
%\IEEEauthorblockN{Guanxiong~Shen, Junqing~Zhang, Alan~Marshall}
%\IEEEauthorblockA{\textit{Department of Electrical Engineering and Electronics} \\
%\textit{University of Liverpool}\\
%Liverpool, L69 3GJ, United Kingdom \\
%Email: \{Guanxiong.Shen, junqing.zhang, alan.marshall\}@liverpool.ac.uk}
%\and
%\IEEEauthorblockN{Linning~Peng}
%\IEEEauthorblockA{\textit{School of Cyber Science and Engineering} \\
%\textit{Southeast University}\\
%No. 2 Sipailou, Nanjing, China\\
%Email: pengln@seu.edu.cn}
%\and
%\IEEEauthorblockN{Xianbin Wang}
%\IEEEauthorblockA{\textit{Department of Electrical and Computer Engineering} \\
%\textit{Western University}\\
%London, Ontario Canada N6A 5B9\\
%Email: xianbin.wang@uwo.ca}
%}

\author{
\IEEEauthorblockN{
Guanxiong~Shen\IEEEauthorrefmark{1},
Junqing~Zhang\IEEEauthorrefmark{1}\IEEEauthorrefmark{4},
Alan~Marshall\IEEEauthorrefmark{1},
Linning~Peng\IEEEauthorrefmark{2},
and Xianbin~Wang\IEEEauthorrefmark{3}
}
\IEEEauthorblockA{
\IEEEauthorrefmark{1}
Department of Electrical Engineering and Electronics, University of Liverpool, Liverpool, L69 3GJ, United Kingdom\\ 
Email: \{Guanxiong.Shen, junqing.zhang, alan.marshall\}@liverpool.ac.uk
}
\IEEEauthorblockA{
\IEEEauthorrefmark{2}
School of Cyber Science and Engineering, Southeast University, No. 2 Sipailou, Nanjing, China\\
Email: pengln@seu.edu.cn
}
\IEEEauthorblockA{
\IEEEauthorrefmark{3}
Department of Electrical and Computer Engineering, 
Western University, London, Ontario, N6A 5B9, Canada\\
Email: xianbin.wang@uwo.ca
}
\IEEEauthorblockA{
\IEEEauthorrefmark{4}
Corresponding Author}
}

%\author{
%	\IEEEauthorblockN{
%		Author 1,
%		Author 2,
%	    Author 3,
%		Author 4,
%	and	Author 5
%	}
%
%}

%\thanks{Manuscript received xxx; revised xxx; accepted xxx. Date of publication xxx; date of current version xxx. 
%The work was supported by Royal Society Research Grants under grant ID RGS\slash R1\slash 191241. 
%The review of this paper was coordinated by xxx.}
%\IEEEcompsocitemizethanks{\IEEEcompsocthanksitem
%G.~Shen, J.~Zhang and A.~Marshall are with the Department of Electrical Engineering and Electronics, University of Liverpool, Liverpool, L69 3GJ, United Kingdom. (email: Guanxiong.Shen@liverpool.ac.uk; junqing.zhang@liverpool.ac.uk; alan.marshall@liverpool.ac.uk)
%\IEEEcompsocthanksitem L. Peng is with the School of Cyber Science and Engineering, Southeast University, No. 2 Sipailou, Nanjing, China and Purple Mountain Laboratories, Nanjing, China (email: pengln@seu.edu.cn).}
%\thanks{Color versions of one or more of the figures in this paper are available online at http://ieeexplore.ieee.org.}
%\thanks{Digital Object Identifier xxx}	
%}

%	{Shell \MakeLowercase{\textit{et al.}}: Bare Demo of IEEEtran.cls for IEEE Journals}

\maketitle

\begin{abstract}
Radio frequency fingerprint identification (RFFI) is an emerging device authentication technique that relies on intrinsic hardware characteristics of wireless devices. We designed an RFFI scheme for Long Range (LoRa) systems based on spectrogram and convolutional neural network (CNN). Specifically, we used spectrogram to represent the fine-grained time-frequency characteristics of LoRa signals. In addition, we revealed that the instantaneous carrier frequency offset (CFO) is drifting, which will result in misclassification and significantly compromise the system stability; we demonstrated CFO compensation is an effective mitigation. Finally, we designed a hybrid classifier that can adjust CNN outputs with the estimated CFO. The mean value of CFO remains relatively stable, hence it can be used to rule out CNN predictions whose estimated CFO falls out of the range. We performed experiments in real wireless environments using 20 LoRa devices under test (DUTs) and a Universal Software Radio Peripheral (USRP) N210 receiver. By comparing with the IQ-based and FFT-based RFFI schemes, our spectrogram-based scheme can reach the best classification accuracy, i.e., 97.61\% for 20 LoRa DUTs.

%we collected training and test data on different days. Our proposed scheme can reach a classification accuracy of 95.52\% for 20 LoRa DUT and remained relatively stable over time, while conventional IQ-based method obtained an accuracy of 85.50\%.
\end{abstract}
	

\section{Introduction}

\IEEEPARstart{T}{he} Internet of things (IoT) applications are blooming with numerous exciting applications such as connected healthcare, smart cities and intelligent industries~\cite{zanella2014internet}. 
Statista estimated there would be 75.44 billion IoT devices by 2025\footnote{https://www.statista.com/statistics/471264/iot-number-of-connected-devices-worldwide/}.
Device authentication is critical to safeguard IoT applications for allowing legitimate users to access the network while preventing malicious users~\cite{xu2015device}. This task is becoming more challenging with the rapid growth of low cost IoT devices. 
%Ericsson predicted that there would be 25 billion IoT devices worldwide by 2025~\cite{IoTpopulation}. 
Conventional authentication schemes rely on software addresses such as Internet Protocol (IP) and/or Media Access Control (MAC) addresses, which are prone to be tampered or forged~\cite{zou2016survey}. 
Once the security credentials are obtained by malicious users, they can masquerade as the 
legitimate users to access the private data or launch fatal attacks on the IoT networks. 

%Recently, there is an emerging non-cryptographic authentication technique known as device fingerprinting~\cite{zeng2010non,xu2015device}. This technique extracts device-specific features and uses them to identify individual devices~\cite{xu2015device}. A variety of features are considered, including network traffic characteristics~\cite{sivanathan2018classifying}, packet inter-arrival time~\cite{luo2019transforming}, clock skew~\cite{jana2009fast}, channel characteristics such as channel state information (CSI)~\cite{liu2017authenticating} and received signal strength (RSS)~\cite{chen2010detecting}, etc. However, network traffic-based schemes require long-term continuous collection to analyze the traffic characteristics, while channel-based methods are dependent on the location or environment. An environment-independent device authentication scheme that can authenticate each individual packet is urgently required.

Radio frequency fingerprint identification (RFFI) is a promising authentication scheme that can identify wireless devices from their emitted transmissions~\cite{xu2015device,riyaz2018deep,zhang2019physical}. Radio frequency fingerprint (RFF) is originated from the hardware imperfections introduced during the manufacturing process, which is inherent to the analog front-end components. 
These imperfections deviate slightly from their nominal specifications hence do not affect the normal communication functions; but we can design advanced algorithms to extract them as a device identifier.
Similar to a biometric fingerprint, RFF is unique and hard to tamper without tremendous efforts. 

RFFI consists of two stages, namely training and classification. During the training stage, an authenticator will collect sufficient wireless packets from devices under test (DUTs), then extract features from the received packets to train a classifier. Various features are considered in previous work including  Hilbert spectrum~\cite{zhang2016specific}, carrier frequency offset (CFO)~\cite{nguyen2011device,hou2014physical,vo2016fingerprinting,hua2018accurate,liu2019real}, inphase and quadrature (IQ) offset~\cite{brik2008wireless}, spectrum\cite{wang2016wireless}, time-frequency statistics~\cite{bihl2016feature}, phase error~\cite{brik2008wireless}, and power amplifier nonlinearity~\cite{polak2011identifying}, etc.
During the classification stage, the authenticator will extract the same type of features from received packets, feed them to the trained classifier and infer the device identity.

Compared with conventional cryptography-based security schemes~\cite{zou2016survey}, one major advantage of RFFI is that it does not impose any additional computational burden and power consumption on the devices to be authenticated~\cite{xu2015device,xie2018optimized}. This is particularly desirable for many IoT applications because most of the end nodes are low-cost with limited computational and energy resources.
For instance, RFFI could be utilized in Long Range (LoRa) networks to relieve the severe battery power constraint of LoRa devices. %Disregard these advantages, RFFI techniques often suffer from low identification reliability due to indistinctive and varying hardware characteristics. 

RFFI can be considered as a multi-class classification problem, hence the most recent development on deep learning could be leveraged~\cite{merchant2018deep,yu2019robust,sankhe2019oracle,sankhe2019no,al2020exposing,ding2018specific,gong2020unsupervised,peng2019deep,pan2019specific,das2018deep}.
Manually extracting handcrafted features requires comprehensive knowledge on the adopted communication technology and protocol. In addition, it is difficult to estimate each individual feature accurately as the hardware imperfections are interrelated~\cite{zhu2013challenges}. Deep learning algorithms can automatically extract features from the received signals and can extract more distinguishable and high-level fingerprints~\cite{yu2019robust}. 
Deep learning-based RFFI systems are built with the latest convolutional neural network (CNN)~\cite{merchant2018deep,yu2019robust,sankhe2019oracle,sankhe2019no,al2020exposing,ding2018specific,gong2020unsupervised,peng2019deep,pan2019specific} or recurrent neural network (RNN) such as long short-term memory (LSTM)~\cite{das2018deep}. Most of them often use the IQ samples as the network input~\cite{merchant2018deep,yu2019robust,sankhe2019oracle,sankhe2019no,al2020exposing}, which may not be the best solution as the signal characteristics is not explicit in the time domain.
The system performance can be improved by transforming IQ samples and obtaining more distinguishable signal representations, such as bispectrum~\cite{ding2018specific,gong2020unsupervised}, differential constellation trace figure~\cite{peng2019deep}, Hilbert-Huang spectrum~\cite{pan2019specific}, error signal~\cite{merchant2018deep}, etc.
During the classification stage, softmax function is often used in neural networks to return a list of probabilities with respect to the classes, indicating the confidence of the predictions. In some cases, the classifier is not confident about its prediction, i.e, confidence score is low and the probabilities of several classes are quite close. However, there is no existing work to leverage this confidence information to calibrate the uncertain predictions.

As a device authentication scheme, RFFI should remain stable~\cite{xu2015device}. Robyns~\textit{et al.}~\cite{robyns2017physical} indicated that the accuracy of their system dropped over time and inferred this was caused by oscillator frequency drift. However, the authors did not provide an in-depth analysis or mitigation methods.
Andrews~\textit{et al.}~\cite{andrews2019extensions} experimentally examined the effect of temperature variation on different analog components, e.g. oscillator, power amplifier, phase locked loop, mixer, etc., and concluded that the oscillator is particularly sensitive to temperature fluctuations. 
%Cekic~\textit{et al.}~\cite{cekic2020robust} indicated that oscillator frequencies may drift substantially over time but did not design experiments to verify it. 
While CFO has been successfully used to identify WiFi devices~\cite{hua2018accurate,liu2019real}, it was also observed that low-cost ZigBee devices have severe CFO variations even within 15 minutes~\cite{peng2018design,yu2019robust}. 
%There is some other research about the impact of supply voltage \cite{danev2009transient,zhuang2018fbsleuth}, their results demonstrate that drift caused by voltage is not serious. 
%CFO compensation is a standard procedure in telecommunication systems, but some RFFI systems take uncompensated IQ samples as the system input~\cite{al2020exposing,sankhe2019oracle,robyns2017physical}.
A comprehensive investigation of the CFO variation on low cost IoT devices and its effect on the RFFI is still missing. 

In this paper, we take LoRa as a case study to investigate the above challenges.
LoRa is a physical layer standard developed by Cycleo and patented by Semtech in 2014~\cite{seller2016low},  which has been widely used for long range IoT applications. 
LoRaWAN, a higher layer protocol for LoRa, relies on cryptography-based schemes for device registration, namely Over The Air Activation (OTAA) and 
Activation by Personalization (ABP)~\cite{LoRaSecurity}, which are prone to be tampered by malicious users. Therefore, the emerging RFFI technique is promising for LoRa device authentication. To the best knowledge of the authors, there are three papers on the LoRa RFFI~\cite{robyns2017physical,das2018deep,jiang2019physical}. 
LoRa employes chirp spread spectrum (CSS) modulation which exhibits time-frequency characteristics, which can be explicitly revealed in spectrogram. 
However, none of them considered the unique modulation techniques of LoRa, which may not be able to reach optimal performance.
In addition, LoRa devices are low cost and usually manufactured with cheap components including oscillators. The effect of CFO variation on the LoRa RFFI  has never been investigated.

This paper designs a CNN-based RFFI system to classify LoRa devices. We aim to answer three questions: (1) Can we employ a signal representation that is unique to LoRa modulation and improve the classification accuracy? (2) How does the CFO variation affect the  RFFI stability and can we mitigate it? (3) Can we leverage the probabilities of the softmax output to further enhance the deep learning-based RFFI? 
We carried out an in-depth investigation and extensive experiments that involved 20 LoRa devices as DUTs and a Universal Software Radio Peripheral (USRP) N210 software defined radio (SDR) platform as the authenticator to answer these questions.
Our contributions are listed as follows.
\begin{itemize}
		\item We experimentally compare three signal representations for LoRa signals, namely IQ samples, Fast Fourier transform (FFT) results and spectrogram. It is found spectrogram can reach the highest accuracy of 96.44$\%$ while the IQ samples and FFT can reach 83.36$\%$ and 87.36$\%$, respectively. In addition to this, the training time of spectrogram-based model (20 minutes) is much shorter than that of IQ/FFT-based model (one hour), which indicates the training cost can be significantly reduced.
		\item We experimentally demonstrate that CFO is unstable and degrades the system performance. 
We established a bespoke setup by connecting a LoRa DUT and USRP with an attenuator to eliminate channel effects. CFO is found to vary in a short time frame but stays relatively stable in long terms. 
CFO compensation is found to be effective in mitigating the performance degradation, which can improve the classification accuracy from 75.59$\%$ to 96.44$\%$ for spectrogram CNN-based scheme. 	
	  \item  We design a hybrid classifier based on the softmax output and CFO to further increase the classification accuracy. The CNN may be uncertain when some devices have very similar hardware characteristics and their softmax output probabilities will be close. As the CFO has long term stability,  we calibrate the output of CNN according to the estimated CFO. The designed hybrid classifier can significantly improve the system performance, namely from 83.36$\%$ to 92.01$\%$ in the best case for the IQ-based RFFI. 
	\end{itemize}

The rest of the paper is organized as follows. Section~\ref{sec:preliminary} briefly introduces the background of LoRa modulation and spectrogram. Then we present the LoRa receiver operations in Section~\ref{sec:loraReceiverOperation}.
The design details of the RFFI system and the CNN architecture are introduced in Section~\ref{sec:system} and Section~\ref{sec:cnn}, respectively. 
In Section~\ref{sec:cfoDiscussion}, we experimentally demonstrate CFO drift and its effect on the stability of RFFI, and in Section~\ref{sec:expWireless} the performance of the proposed RFFI systems is thoroughly evaluated in a real wireless environment. The paper is finally concluded in Section~\ref{sec:conclusion}.
	
\section{Preliminary}\label{sec:preliminary}
\subsection{LoRa Modulation Technique}\label{sec:LoRaIntroduction}
LoRa employs CSS modulation which uses linear chirps for communications. The linear chirps are also known as linear frequency modulation (LFM) signals, of which the frequency increases or decreases linearly with time. The information in each symbol is encoded to the initial phase of the chirp. A standard basic chirp $c(t)$ of the RF band can be mathematically expressed as 
\begin{equation}
	c^{rf}(t) =A e^{ j(-\pi Bt + \pi\frac{B}{T}t^2+ 2\pi f_ct)} = u(t) e^{j2\pi f_ct},
\end{equation}
where $f_c$ is the carrier frequency, $B$ is the bandwidth, $T$ is the symbol duration, and $A$ is the amplitude.
The baseband signal can be given as
\begin{align}
	u(t) = Ae^{ j(-\pi Bt + \pi\frac{B}{T}t^2)} =  Ae^{j\phi(t)},
\end{align}
where $\phi(t)$ denotes the phase of the baseband chirp signal. The symbol duration $T$ is given as
\begin{equation}
T = \frac{2^{SF}}{B},
\end{equation}
where $SF$ is the spreading factor ranging from 7 to 12.  The instantaneous frequency of $u(t)$ is defined as
\begin{equation}
f(t) = \frac{1}{2\pi} \frac{d\phi(t)}{dt} = -\frac{B}{2} + \frac{B}{T}t.
\label{eq:inst_freq}
\end{equation}

\begin{figure}[!t]
	\centering
	\subfloat[]{\includegraphics[width=1.7in]{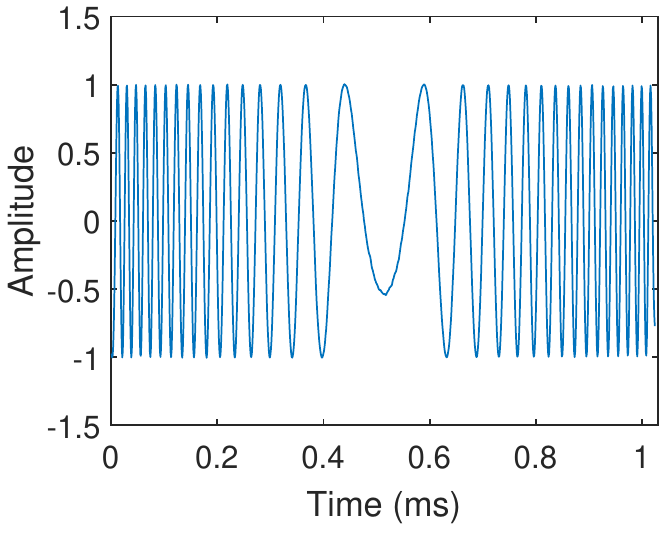}
		\label{fig:oneChirp}}
	\subfloat[]{\includegraphics[width=1.7in]{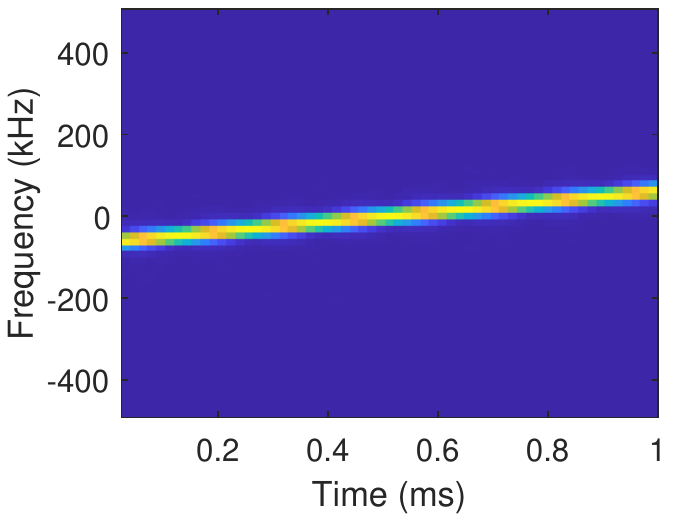}
    \label{fig:oneChirpSpectrogram}}
	\caption{LoRa preamble. (a) Time domain representation of one preamble (I branch). (b) Spectrogram of one preamble. }
	\label{fig:STFT}
\end{figure}

The LoRa standard specifies several basic chirps called preambles in a packet~\cite{lorawanRegionalParameters}. The preamble is the same in every packet and for any LoRa device type. Fig.~\ref{fig:oneChirp} shows the time-domain baseband signal (I branch) of one preamble in a LoRa packet.

\subsection{Short-Time Fourier Transform and Spectrogram}\label{sec:STFT}
Short-time Fourier transform (STFT) is a well-known time-frequency analysis algorithm which has been extensively utilized to analyze non-stationary signals, including LoRa signals. STFT divides a long signal into short segments and then performs Fourier transform separately on each segment. The discrete-time STFT can be mathematically given as 
\begin{equation} \label{equ.stft}
STFT(m,f) = \sum_{n=-\infty}^{\infty} s[n]w[n-mR]e^{-j2\pi fnT_s} ,
\end{equation}
where $s[n]$ is the signal to be analyzed, $w[n]$ is the window function of length $M$, $m$ is the column index of the matrix and $R$ is the hop size. The spectrogram can be given as
\begin{equation} \label{equ:spectrogram}
Spectrogram(m,f) = \left |  STFT(m,f)\right |^{2},
\end{equation}
where $|\cdot|$ returns the amplitude.
Fig.~\ref{fig:oneChirpSpectrogram} is the spectrogram of one LoRa preamble.
Spectrogram can efficiently represent how the instantaneous frequency changes over time, as well as some signal parameters such as bandwidth $B$ and symbol duration $T$.

\section{LoRa Receiver Operation}\label{sec:loraReceiverOperation}
	
\subsection{Signal Reception}\label{sec:downconversion}
The LoRa signal is first received by the receiver antenna, i.e., $r^{rf}(t)$. Then it is down-converted to the baseband by the mixer. The received baseband signal is sampled by an analog-to-digital converter (ADC) to obtain the digital baseband signal $r[nT_s]$, which can be mathematically expressed as
\begin{align}
r[nT_s] =& r^{rf}[nT_s]e^{-j2\pi f^{rx}_c nT_s} \nonumber\\
=& u'[nT_s] e^{j2\pi f^{tx}_cnT_s} e^{-j2\pi f^{rx}_cnT_s}\nonumber\\
=& u'[nT_s]e^{j2\pi \Delta fnT_s},
\label{equ:downConversion}
\end{align}
where $u'[nT_s]$ is the received baseband signal, $T_s$ is the sampling interval, $f^{tx}_c$ and $f^{rx}_c$ are the carrier frequencies of the transmitter and receiver, respectively, and $\Delta f = f^{tx}_c - f^{rx}_c$ is the CFO between them. 
For the simplicity of notations, $T_s$ is omitted. The digital baseband signal can be rewritten as
\begin{equation} 
r[n] = u'[n]e^{j2\pi \Delta fnT_s}.
\end{equation}

\subsection{CFO Estimation and Compensation}\label{sec:cfoEstimation}
%As shown in (\ref{equ:downConversion}), there is usually an inevitable offset between $f^{tx}_c$ and $f^{rx}_c$ due to the hardware imperfection, which is defined as the CFO. 
%In addition, the movement of transmitter or receiver creates a Doppler shift which leads to a CFO as well, but it is relatively minor in low-speed scenario compared to the CFO resulting from the oscillator mismatch, which is not considered in this paper.
In this section, we will introduce the CFO estimation and compensation for LoRa signals. 

%Many previous studies constructed CFO-based RFF identification systems~\cite{hou2014physical,hua2018accurate}. However, we experimentally found that CFO is not a stable feature for IoT devices which leads to system performance degradation over time. CFO compensation is effective in solving this problem so it is an indispensable step in deep learning-based RFF system. Detailed discussion and experimental evidence about the drift of CFO are presented in Section~\ref{sec:cfoDiscussion}.\red{DO NOT put this part here. You don't have these results in this section.}

\subsubsection{Coarse CFO Estimation}\label{sec:coarseEst}
The ideal instantaneous frequency of the baseband basic chirp, $f_{ideal}[n]$, increases linearly from $-\frac{B}{2}$ to $\frac{B}{2}$. 
However, there is an inevitable frequency offset, $\Delta f$, in the received baseband signal $r[n]$. Its instantaneous frequency $f[n]$ thus becomes
\begin{equation} \label{}
f[n] = -\frac{B}{2} + \Delta f +\frac{B}{T}nT_s .
\end{equation}

Thanks to the linearity of $f[n]$, the CFO can be roughly estimated by calculating the mean value of  $f[n]$ of the received preambles. The estimated CFO $\widehat f_{coarse}$ is given as
\begin{equation} \label{equ:coarseEst}
\Delta \widehat f_{coarse} = \frac{1}{L} \sum_{n=0}^{L-1} f[n],
\end{equation}
where $L$ is the symbol length, defined as
\begin{equation} 
L = \frac{T}{T_s} = \frac{2^{SF}}{B\cdot T_s}.
\end{equation}

The received signal can be coarsely compensated by the estimated frequency offset, given as
\begin{equation} \label{equ:coarseCompensation}
{r}'[n] =  r[n]\cdot e^{-j2\pi \Delta \widehat{f}_{coarse} nT_s} .
\end{equation}

%\begin{equation} \label{}
%\Delta \widehat f_{coarse} = r[n] = u[n]e^{j2\pi \Delta fnT_s} = \frac{1}{2\pi} \frac{d\phi(t)}{dt}
%\end{equation}

\subsubsection{Fine CFO Estimation}
There will be residual frequency offset after the above coarse frequency compensation, hence we further employ a fine estimation algorithm. The residual offset  can be estimated based on the repeating property of preambles, given as
%The CFO can be roughly estimated by the approach proposed in the previous subsection. However, in such an approach we cannot obtain the accurate CFO, there is usually an estimation error of tens of hertz. In this subsection, We aim to more precisely estimate the offset which may further improve the system performance.
%After the process in (\ref{equ:coarseCompensation}), there still exists a residual offset since the coarse estimation (\ref{equ:coarseEst}) is not accurate enough. 
\begin{equation} \label{equ:fineEst}
\Delta \widehat f_{fine} = -\frac{1}{2\pi \cdot T_s L} \cdot \angle \Big(\sum_{n=0}^{L-1}{r}'[n]\cdot r'^*[n+L] \Big) ,
\end{equation}
where $\angle \cdot$ returns the angle of the variable and $(\cdot)^*$ denotes conjugation. 
The received signal can be further finely compensated as 
\begin{equation} \label{equ:fineCompensation}
{r}''[n] =  r'[n] \cdot e^{-j2\pi \Delta \widehat f_{fine} nT_s}.
\end{equation}

As the phase can only be resolved in $[-\pi,\pi]$, the range of CFO that can be estimated by (\ref{equ:fineEst}) is 
\begin{equation}
 |\Delta \widehat f_{fine}| < \frac{\pi}{2 \pi \cdot T_s L} = \frac{B}{2^{SF+1}}.
\end{equation} 
When the LoRa transmission is configured with $SF$=7 and $B$=125 kHz, the estimation capability is within $\pm$488.3 Hz.
It is common that the frequency drift of the oscillator of LoRa devices is $\pm$10 ppm~\cite{frequencyTolerence}, approximately 8.68 kHz for an 868~MHz carrier frequency, which is much higher than 488.3~Hz. Hence, the coarse estimation should be employed to  
limit the residual offset before the fine estimation.

After the coarse and fine CFO estimation, the overall estimated CFO, $\Delta \widehat f$, can be represented as
\begin{equation}
 \Delta \widehat f = \Delta \widehat f_{coarse} + \Delta \widehat f_{fine}.
\end{equation}

\section{RFFI System}\label{sec:system}
\begin{figure}[!t]
	\begin{center}
		\includegraphics[width = 3.4in]{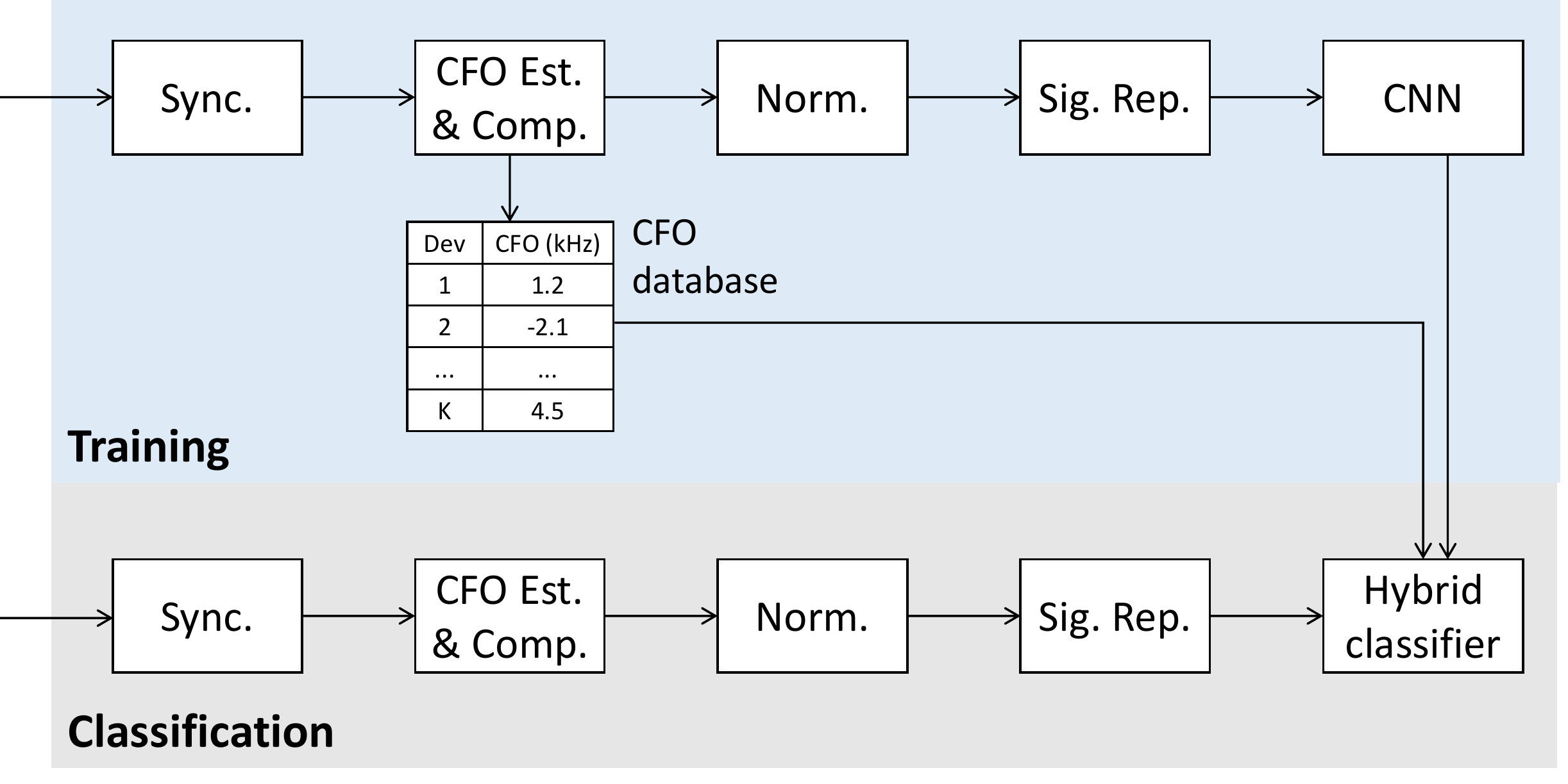}
		\caption{A CNN-based RFFI scheme. CFO compensation is adopted.}
		\label{fig:systemArchitecture}
	\end{center}	
\end{figure}

The architecture of the proposed RFFI system is shown in Fig.~\ref{fig:systemArchitecture}. This section will introduce each step in detail.

%The received signal is first downconverted to the baseband, synchronization, CFO estimation and compensation are then performed. Next, the processed baseband signal is normalized and converted to a spectrogram by STFT. The obtained spectrogram which can be regarded as an image is directly fed into the CNN for classification. At the training stage, a CFO reference database is generated to form a hybrid classifier together with the CNN, which further improves the system performance by calibrating softmax outputs of the CNN classifier. 

\subsection{Synchronization and CFO Compensation}
Synchronization detects the signal arrival and locates the packet relying on the repeated preambles, which is a standard process in the communication system. Interested readers please refer to the work in~\cite{robyns2018multi} for detailed information.

CFO estimation and compensation are standard procedures in wireless communication systems as well. However, some previous studies did not perform these steps as they used the raw IQ samples directly. Some work also used CFO as one of the RFF features~\cite{nguyen2011device,hou2014physical,vo2016fingerprinting,hua2018accurate,liu2019real}. However, studies also revealed CFO would cause performance degradation~\cite{robyns2017physical}.
The effect of CFO on the RFFI for low-cost IoT devices is not experimentally investigated.

We adopted the CFO estimation and compensation algorithms introduced in Section~\ref{sec:cfoEstimation}. During the training, a CFO database is generated, which contains the estimated CFO of each DUT. This CFO database will be used for the hybrid classifier which will be introduced in Section~\ref{sec:hybrid}.

%As introduced in Section~\ref{sec:cfoEstimation}, there is an inevitable CFO in the received signal $r[n]$. Even though CFO compensation is necessary to ensure the normal operation of the communication system, many previous RFFI studies use the uncompensated signal as system inputs. Because taking the original signal as the input to the system does not lose any information.
%
%We experimentally found that CFO of low-cost IoT devices is not stable. The instability of CFO will reduce the robustness of RFFI systems, i.e, the system performance drops over time. Detailed experimental results and discussion can be found in Section~\ref{sec:cfoDiscussion} and Section~\ref{sec:expWireless}. Hence we reckon that CFO compensation is indispensable for RFFI system. Detailed algorithms for CFO estimation and compensation can be found in Section.~\ref{sec:cfoEstimation}.

\subsection{Normalization}\label{sec:normalization}
RFFI systems are not expected to differentiate devices by power differences because signal power is susceptible to distance. Normalization has been a standard process in RFFI. 
%All the samples of the received packet are divided by the root mean square (RMS) of amplitude to normalize each sample to unit power, 
The normalized signal $s[n]$ can be given as
\begin{equation}
s[n] = \frac{r''[n]}{x_{rms}},
\end{equation}
where $x_{rms}$ is the root mean square of the amplitude of $r''[n]$.

\subsection{Signal Representation}\label{sec:signal_representation}
%Excellent signal representations can significantly improve the performance of deep learning-based RFFI system, how to find the best representations has become a key challenge. 
Signal representation employs signal processing algorithms to reveal the underlying signal characteristics, which can be better learned by the classifier. This paper only uses the preamble part to prevent the deep learning model learning protocol-specific and data-specific knowledge.

\subsubsection{IQ Samples}
IQ samples represent the time-domain signals which are captured from the receiver chain directly.
Some previous work aims to design protocol-agnostic RFFI systems without considering the physical modulation schemes so they employ IQ samples as system inputs~\cite{merchant2018deep,yu2019robust,sankhe2019oracle,sankhe2019no,al2020exposing}.

\subsubsection{FFT Results}
FFT converts the time-domain signal to the frequency domain. Features that are not obvious in the time domain may be easily observed in the frequency domain. The FFT coefficients are readily available in WiFi OFDM systems~\cite{al2020exposing}.

\subsubsection{Spectrogram}
%As introduced in Section~\ref{sec:STFT}, STFT has been widely used to analyze non-stationary signals, including chirps employed by LoRa. It is also an efficient approach to represent LoRa symbols, as shown in Fig~\ref{fig:oneChirpSpectrogram}.

The spectrogram can be a better signal representation for LoRa signals since it converts the time domain IQ samples to the time-frequency domain characteristics, which not only provides information in the frequency domain but also reveals how it changes over time. Logarithmic compression of magnitudes was found to be effective in improving the performance and has been a standard strategy in preprocessing spectrograms~\cite{choi2018comparison}, which is also used in this paper.

\subsection{Convolutional Neural Network}
CNN has attracted many research interests from both academia and industry thanks to its excellent performance in image recognition and computer vision. It can find patterns in the data automatically which eliminates the need for manual feature extraction. CNN is usually composed of convolutional layers, fully connected layers, as well as some pooling layers that reduce the number of parameters to prevent overfitting. The convolutional and pooling layers act as a feature extractor that directly extracts features from the input data. The extracted high-level features are then fed into the fully connected layers for classification.

% CNNs can be regarded as the regularized version of multilayer perceptrons (MLP), which can avoid over-fitting problem to a certain extent. Kernels are small matrices that are often used for blurring, sharpening and edge detection in conventional image processing tasks, which are commonly designed manually by experts. In contrast to primitive methods, CNNs are able to learn the parameter of kernels automatically with enough training, thus can extract features (texture, pattern) from the input images. For 1D inputs, such as time-series signal, 1D convolution is usually applied instead of 2D convolution to capture hidden features.\junqing{discussion}

In classification problems, softmax function is usually used at the last layer of CNN to map its outputs to a list of probabilities $\mathbf{S} = (S_1, S_2, ..., S_K)$ over all the predicted classes. ${S}_k$ is the predicted probability of the $k$-th class, which can be mathematically expressed as
\begin{equation} 
\label{eq:softmax}
\begin{aligned}
  {S}_k= \sigma(\mathbf z)_k = \frac{e^{z_k}}{\sum_{j=1}^{K}e^{z_j}} \quad \text{for} \quad  k &= 1,2,...,K,
 \end{aligned} 
\end{equation}
where $K$ is the number of classes, and $\mathbf z = (z_1,z_2,...,z_K)$ is the output of the layer before softmax activation. The most common way to make a prediction is to select the class with the highest probability as the final predicted label.

The CNN architectures used in this paper will be elaborated in Section~\ref{sec:cnn}.

% and the spectrogram obtained by STFT is then directly fed into the CNN for classification.

%\begin{figure}[t!]
%	\begin{center}
%		\includegraphics[width = 3.4in]{CNNArchitecture.pdf}
%		\caption{Convolutional neural network architecture.}
%		\label{CNNArchitecture}
%	\end{center}
%\end{figure}

\subsection{Hybrid Classifier}\label{sec:hybrid}
%Although CNN has already been able to distinguish among different devices from the compensated spectrograms without any CFO information, it is still desired that CFO can contribute to the identification procedures because it is an important hardware-originated feature and has been widely adopted in previous research. Our proposed algorithm can adjust the output of CNN based on the estimated CFO, in this way, the CFO feature is successfully considered.

%It is not proper to cancel the process of CFO compensation to enable CNN to learn the CFO feature. Because in that case, the CNN will treat CFO as a dominant feature but it is not stable enough. A slight change of CFO will make the identification system less effective, relevant experimental results can be found in Section~\ref{sec:expWired}. However, according to the results shown in Fig.~\ref{fig:cfoShortTime} and Fig.~\ref{fig:cfoLongTime}, CFO will not change too much over time. Hence it is possible to exploit CFO feature as a reference to judge whether the prediction of CNN is correct. In other words, CFO is not suitable as a feature for classification but effective in excluding some impossible outputs derived by CNN.

CNN cannot perfectly distinguish devices whose hardware characteristics are quite similar, particularly when they are from the same manufacturer. 
Then, the output probabilities of these classes are close to each other, e.g.,  $S_1$=0.51 and $S_2$=0.49. 
In this case, simply selecting the device with the highest probability is likely to cause misclassification.

As we will demonstrate later in Section~\ref{sec:cfoDiscussion} and Fig.~\ref{fig:cfoLongTime}, the mean values of CFO among different days remained relatively stable. Hence, it inspires us to use the estimated CFO to calibrate the CNN predictions.

We propose a hybrid classifier to exclude unreliable predictions derived by the CNN classifier, which is described in Algorithm~\ref{alg:hybrid}.
%Suppose there are $K$ candidate devices, thus the output vector $\mathbf{S}$ of CNN should have $K$ elements, each element representing the probability of the corresponding device;
%	\item [2)]
%	The highest $N$ elements are selected from $\mathbf{S}$ to form a new vector $\mathbf{P} = (P_1, P_2, ..., P_N)$. The received packet should come from one of these $N$ devices;
We first create a reference CFO database for all the $K$ devices during the training stage, namely, $\{\Delta \widehat{f}_k$\}.
Then, for each DUT during the classifications stage, we will estimate its CFO, $\{\Delta \widehat{f}_{DUT}$\}, and compare it with the CFO database. The operation can be formulated as a hypothesis test
%When the difference between $\Delta \widehat{f}_{DUT}$ and $\Delta \widehat{f}_k$ is larger than a threshold $\lambda$, 
\begin{equation} 
	\left |\Delta \widehat{f}_{DUT} - \Delta \widehat{f}_k \right | \underset{\mathcal H_0}{\overset{\mathcal H_1}{\gtrless}} \lambda,
	\end{equation}
where $\lambda$ is the predefined threshold based on the range of CFO variations. Hypothesis $\mathcal H_1$ means that the packet is impossible to be sent from the $k$-th device due to the large difference between $\Delta \widehat{f}_{DUT}$ and the reference $\Delta \widehat{f}_k$. 
When this occurs, the probability of $k$-th class, $S_k$, is set to zero.
In contrast, hypothesis $\mathcal H_0$ means the prediction of CNN is correct, thus  $S_k$ maintains the original value.
After such calibration, the device with the highest probability in $\mathbf{S}$ is selected as the final predicted label.
\begin{algorithm}[!t]
	\caption{Hybrid Classifier}
	\begin{algorithmic}[1]
		\INPUT{$\mathbf{S}$, The softmax output which denotes the probability of each device;}
		\INPUT{$\Delta \widehat{f}_{DUT}$, The estimated CFO of the DUT;}
		\INPUT{$\Delta\widehat{f}_k$, The reference CFO of the $k$-th device stored in the database;}
		\INPUT{$\lambda$, The CFO threshold.}
		\OUTPUT{$l$, The eventually predicted label.}
		
		\For{$k$ = 1 \textbf{to} $K$}
		\If{$ \left| \Delta \widehat{f}_{DUT} - \Delta\widehat{f}_k \right| > \lambda $}
		\State{$S_k = 0$}
		\Else
		\State{$S_k = S_k$}
		\EndIf
		\EndFor
		\State{Select the device with the highest probability in $\mathbf{S}$ as the predicted label.}
	\end{algorithmic}
	\label{alg:hybrid}
\end{algorithm}

%Using such a hybrid classifier, CFO can help CNN exclude impossible predictions thus further improve the performance of the proposed RFFI system. 
%\xiaowei{can you come up with some explanation to this success? }

\section{CNN Architecture}\label{sec:cnn}

\subsection{Spectrogram-based CNN}

\begin{figure}[!t]
	\centering
	\subfloat[]{\includegraphics[width=3.4in]{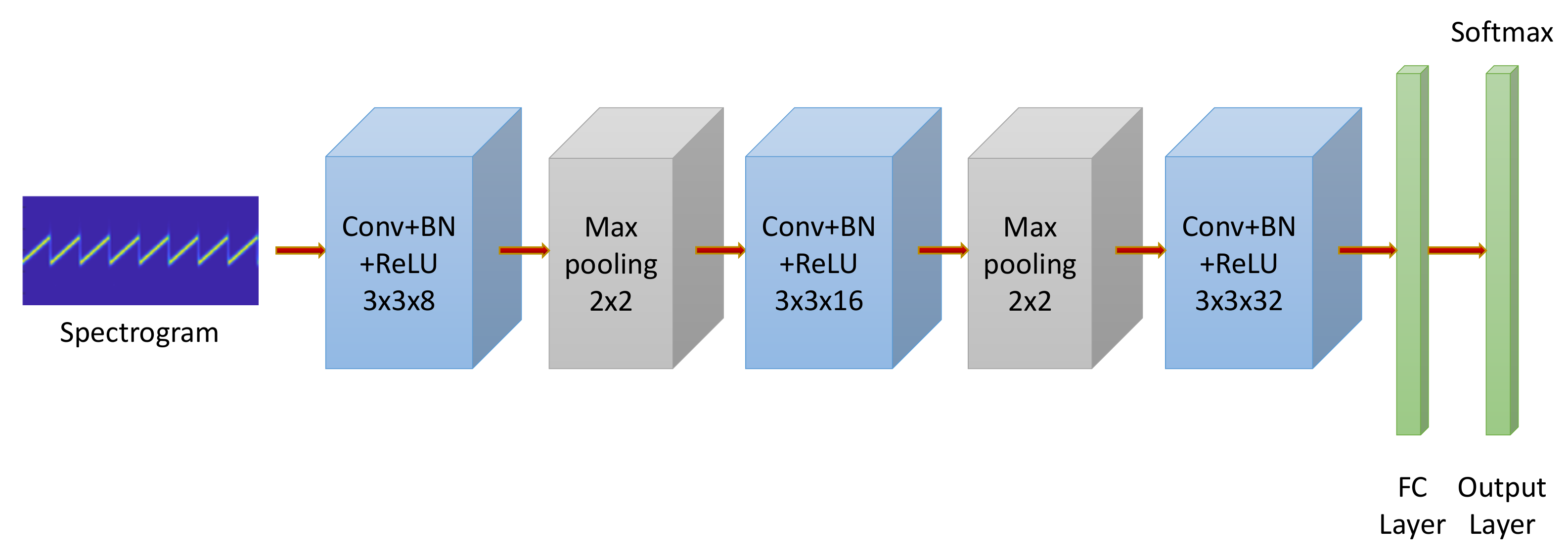}
		\label{fig:stftModel}}\\
		\subfloat[]{\includegraphics[width=3.4in]{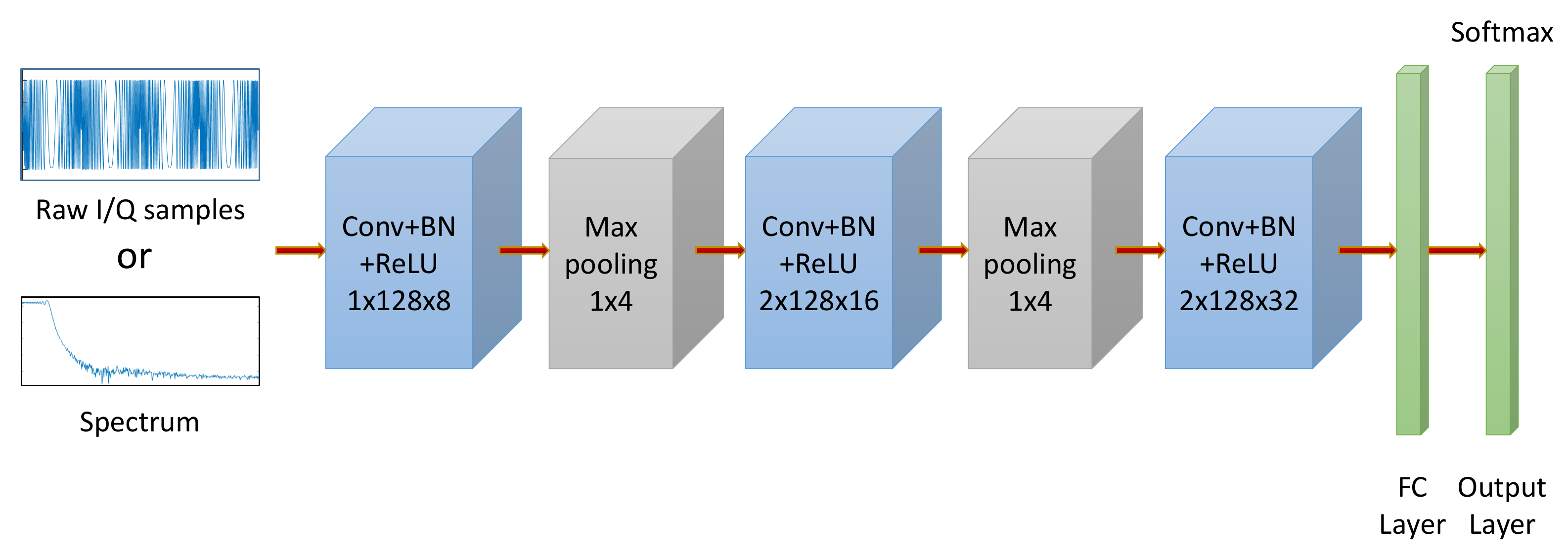}
		\label{fig:baselineModel}}
	\caption{CNN architectures. (a) Spectrogram-based model. (b) IQ/FFT-based model.}
	\label{fig:cnn}
\end{figure}

The architecture of spectrogram-based CNN model is illustrated in Fig.~\ref{fig:stftModel}. It consists of three convolutional layers of 8, 16, and 32 3$\times$3 filters, respectively. Each convolutional layer is followed by a batch normalization layer, the rectified linear unit (ReLU) activation and a 2$\times$2 max pooling layer with stride 2. The output of the last convolutional layer is then fed into the fully connected layer for classification. The softmax function is used in the last layer of the neural network to output the probabilities for different classes.

Adam is selected as the optimizer for the training process. The initial training rate is set to 0.0003 and drops every 10 epochs with a drop factor of 0.3. The mini-batch size is set to 32. 

The spectrogram used in this paper is generated with the rectangular window, with a window length $M$ of 256 and hop size $R$ of 128. Our network is designed for this specific parameter setting.

\subsection{IQ/FFT-based CNN}\label{sec:baseline}

%In order to fairly evaluate the superiority of the spectrogram-based model, 
We also establish another CNN model which can input the complex IQ samples or FFT results, i.e, consider the real and imaginary part of the complex signal as two independent dimensions~\cite{sankhe2019oracle}. It is worth noting that IQ samples and corresponding FFT results are complex vectors of the same length. Hence exactly the same CNN can be used for both IQ and FFT data.

The architecture of the IQ/FFT-based CNN is shown in Fig.~\ref{fig:baselineModel}, which also consists of three convolutional layers and one fully connected layer.
The three convolutional layers are composed of 8, 16, and 32 filters, respectively, and the filter sizes are  $1\times 128$, $2\times 128$, and $2\times 128$, respectively. Each convolutional layer is followed by a batch normalization layer and ReLU function is selected as the activation function. There are two max pooling layers of size $1\times 4$ following the first and second convolutional layer, respectively. The output of the third convolutional layer is fed into a fully connected layer for classification, and the softmax activation function is selected to output the probability of each class. 

To make a fair comparison, the IQ/FFT-based and the spectrogram-based CNNs are deliberately designed with similar network architectures and trained under the same settings such as initial learning rate and mini-batch size.
Both of them are implemented using the Matlab Deep Learning Toolbox\footnote{https://mathworks.com/products/deep-learning.html}.
%Therefore, The difference in performance is not due to the difference in network complexity.

\section{Experimental Results of CFO Drift} \label{sec:cfoDiscussion}
%Verify The Instability of CFO / The Indispensability of CFO Compensation

The RF fingerprints must be time-invariant in the presence of environmental changes as they represent the user identities.
In this section, we experimentally demonstrated that the CFO of LoRa devices drifts over time and CFO compensation is an essential procedure to mitigate performance degradation.

\subsection{Experimental Setup}\label{sec:setup}
\begin{figure}[!t]
	\centering
	\subfloat[]{\includegraphics[width=1.65in]{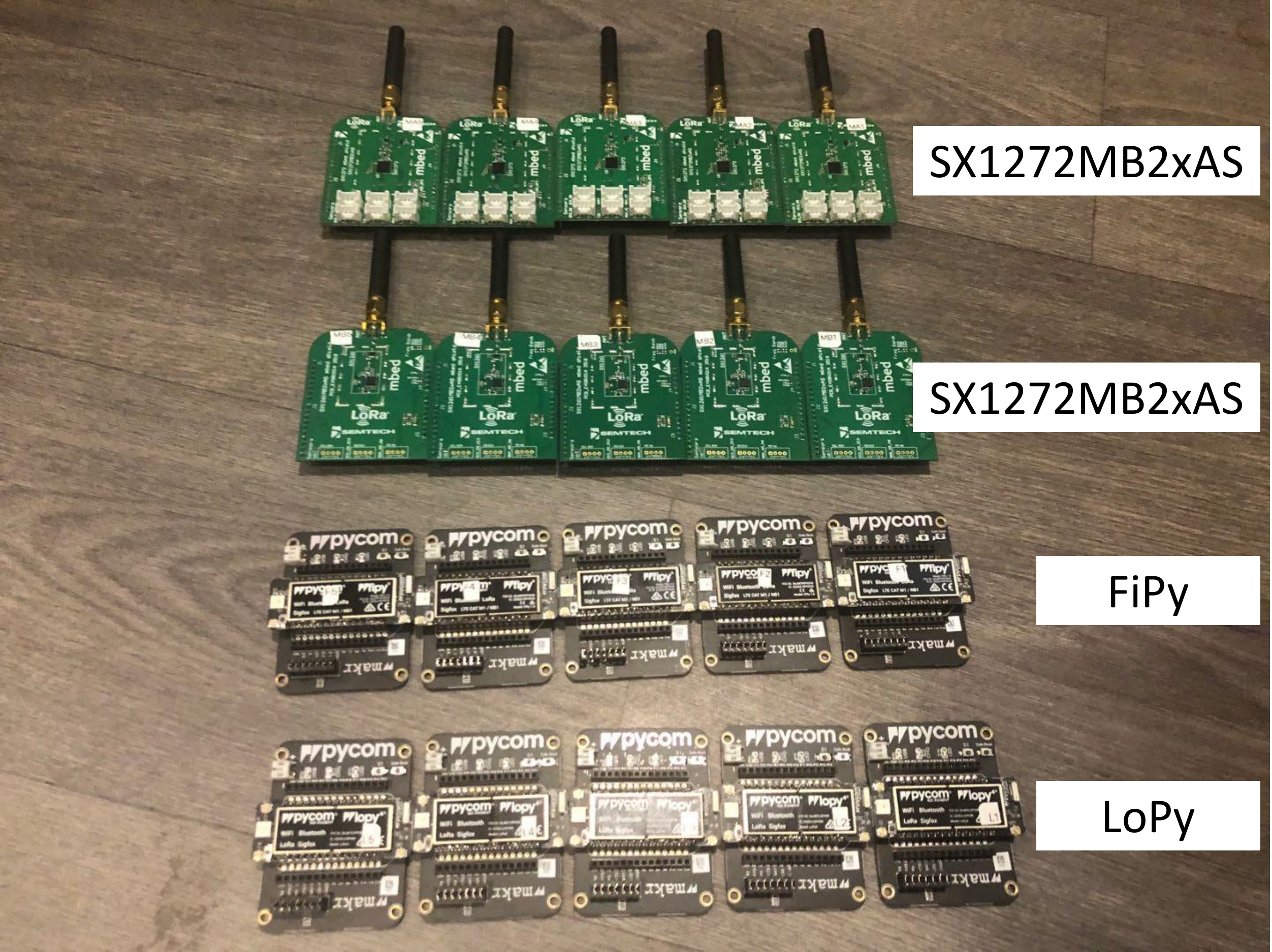}
		\label{fig:expDevices}}
	\subfloat[]{\includegraphics[width=1.65in]{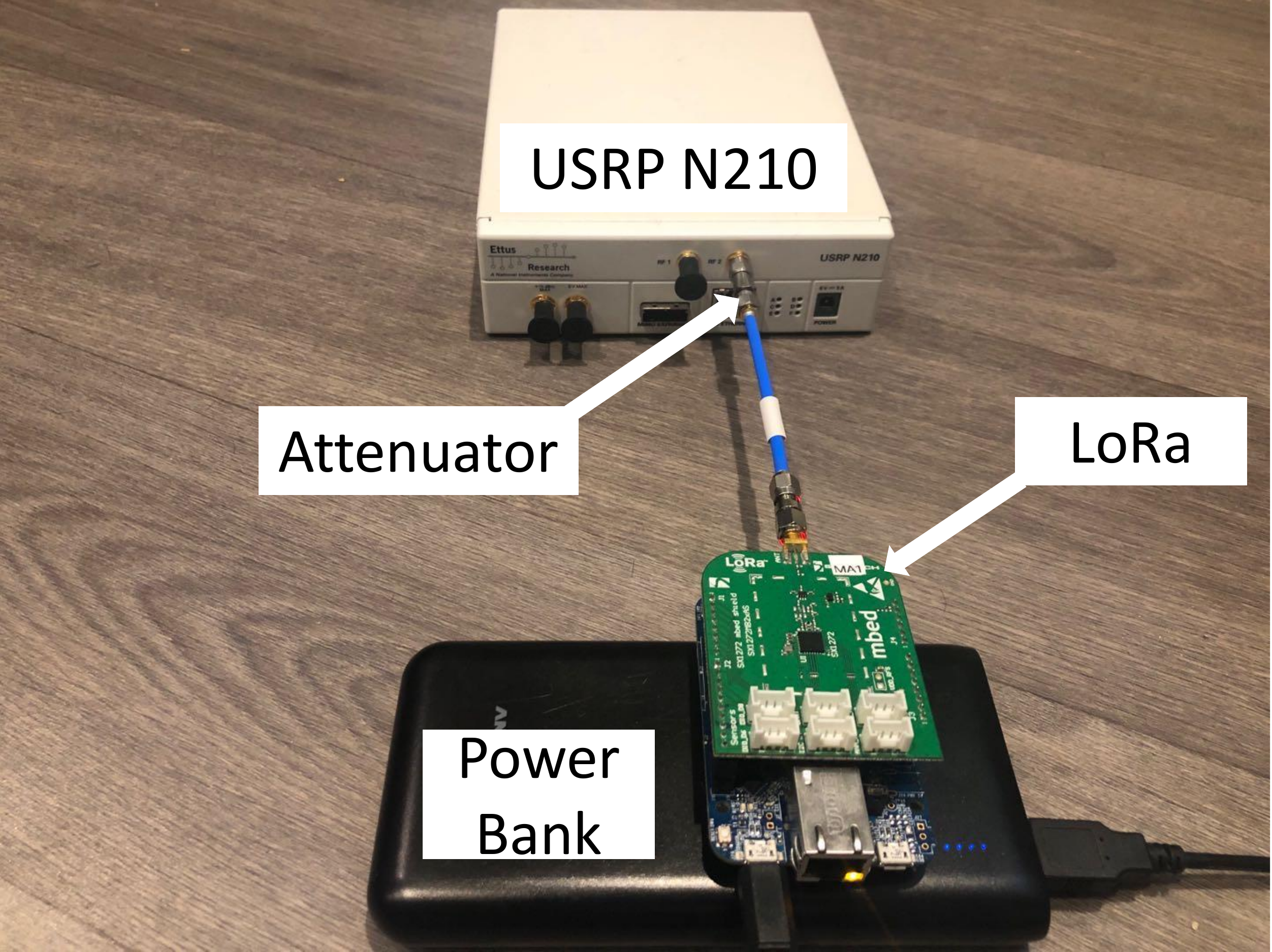}
		\label{fig:wiredEnvironment}}
	\caption{Experimental devices and setup. (a) LoRa DUTs. (b) The LoRa transmitter and USRP receiver connected by a 40~dB attenuator.}
	\label{fig:expEnvironments_wired}
\end{figure}

We used ten LoRa devices of two models, namely five SX1272MB2xAS mbed shields and five SX126xMB2xAS mbed shields, as listed in Table~\ref{tab:devInformation} and shown in Fig.~\ref{fig:expDevices}. %\junqing{Is Table I necessary if we do not mention devices from the same vendor may have similar features? (The red rectangular areas in the confusion matrix)}. 
All the LoRa devices were configured with $SF$ = 7, bandwidth $B$ = 125~kHz, and carrier frequency $f_c$ = 868.1~MHz.
The receiver was a USRP N210 SDR and configured with carrier frequency $f_c$ = 868.1~MHz and 1 MS/s sampling rate. We used the Communications Toolbox Support Package for USRP Radio of Matlab\footnote{https://mathworks.com/help/supportpkg/usrpradio/} to control the USRP and access IQ samples from it.
In order to eliminate channel effects and focus on CFO variations, we created a bespoke setup by connecting the LoRa DUT and USRP N210 receiver by a 40 dB attenuator, as shown in Fig.~\ref{fig:wiredEnvironment}. 
\begin{table}[!t]
  \centering
  \caption{LoRa DUTs.}
    \begin{tabular}{|l|l|l|}
    \hline
    DUT Index & Model  & Chipset \bigstrut\\
    \hline
    1 - 5   & SX1272MB2xAS mbed shield\footnote{https://os.mbed.com/components/SX1272MB2xAS/} & SX1272 \bigstrut\\
    \hline
    6 - 10  & SX126xMB2xAS mbed shield\footnote{https://os.mbed.com/components/SX126xMB2xAS/} & SX1261 \bigstrut\\
    \hline
    11 - 15 & Pycom FiPy\footnote{https://pycom.io/product/fipy/} & SX2172 \bigstrut\\
    \hline
    16 - 20 & Pycom LoPy\footnote{https://pycom.io/product/lopy4/} & SX1276 \bigstrut\\
    \hline
    \end{tabular}%
  \label{tab:devInformation}%
\end{table}%

The data collection for each device lasted for about one hour and was repeated for four days. The transmission interval was set to 1 second and 3,000 packets were collected in about one hour, considering the packet duration and processing time. We named the four datasets as Day 1, Day 2, Day 3 and Day 4 dataset.

%In such case, the channel condition remained the same on different days and therefore the performance deterioration is not due to the channel changes but to changes of the device itself. 

%1,000 packets collected on Day1 were used to train the CNN, of which 85$\%$ is the training set and the rest 15$\%$ is the validation set. In addition, another 1,000 packets were collected on Day1 to form the Day1 test set. On Day2, Day3 and Day4, we collected 1,000 packets each day and named as Day 2, Day 3 and Day 4 test set respectively.

\subsection{CFO Drift}\label{sec:stabilityCFO}
The CFO drift is demonstrated from two aspects, namely short-time and long-time variation. The short-time variation refers to the CFO of devices changed rapidly within a short time after they are powered on while the long-time variation shows that the average CFO drifts within four days but remains relatively stable.

\subsubsection{Short-time Variation}
\begin{figure}[!t]
	\centering
	\subfloat[]{\includegraphics[width=1.65in]{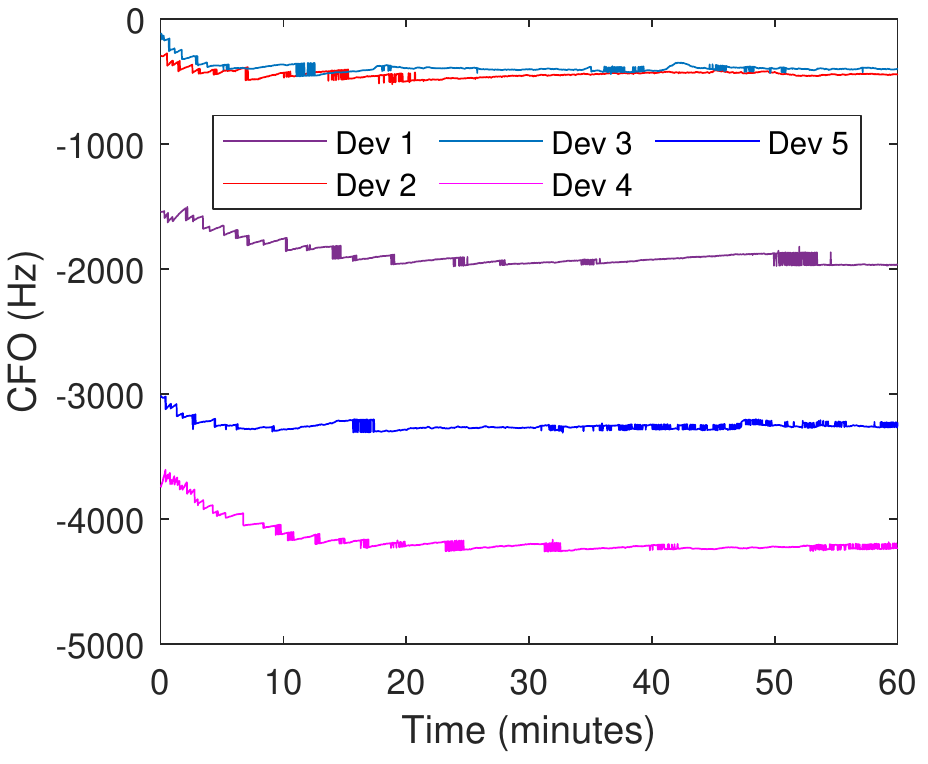}
		\label{fig:cfoshortTime1_5}}
	\subfloat[]{\includegraphics[width=1.65in]{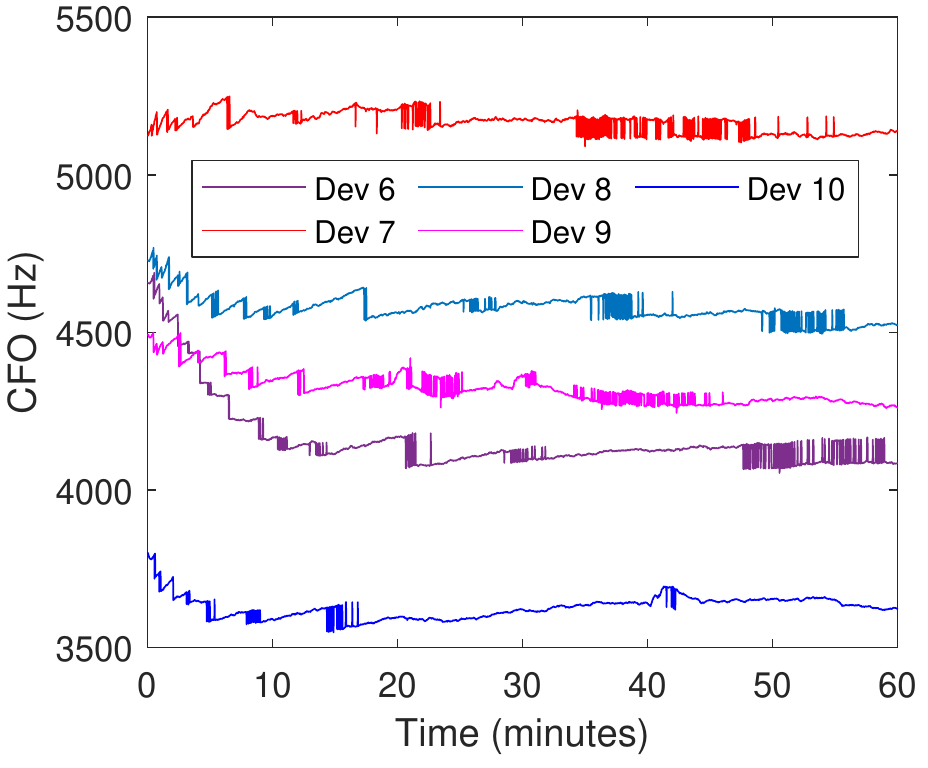}
		\label{fig:cfoshortTime6-10}}\\
	\caption{CFO variations within one hour. (a) Dev 1-5. (b) Dev 6-10.}
	\label{fig:cfoShortTime}
\end{figure}

We analyzed the Day 1 dataset as an example to observe how the CFO changes within one hour. The CFO of each packet was estimated using the algorithm introduced in Section~\ref{sec:cfoEstimation}. As shown in Fig.~\ref{fig:cfoShortTime}, the CFO of each device decreased over the first 20 minutes and then remained relatively constant. This is reasonable because the temperature gradually increases after the device is powered on (self-heating) and the oscillator is sensitive to temperature variations~\cite{andrews2019extensions}. 

%The short-time instability of CFO means that when the LoRa device is just powered on, the CFO-based RFF system probably cannot identify it correctly which is unacceptable.

\subsubsection{Long-time Variation}
\begin{figure}[!t]
	\centering
	\subfloat[]{\includegraphics[width=1.7in]{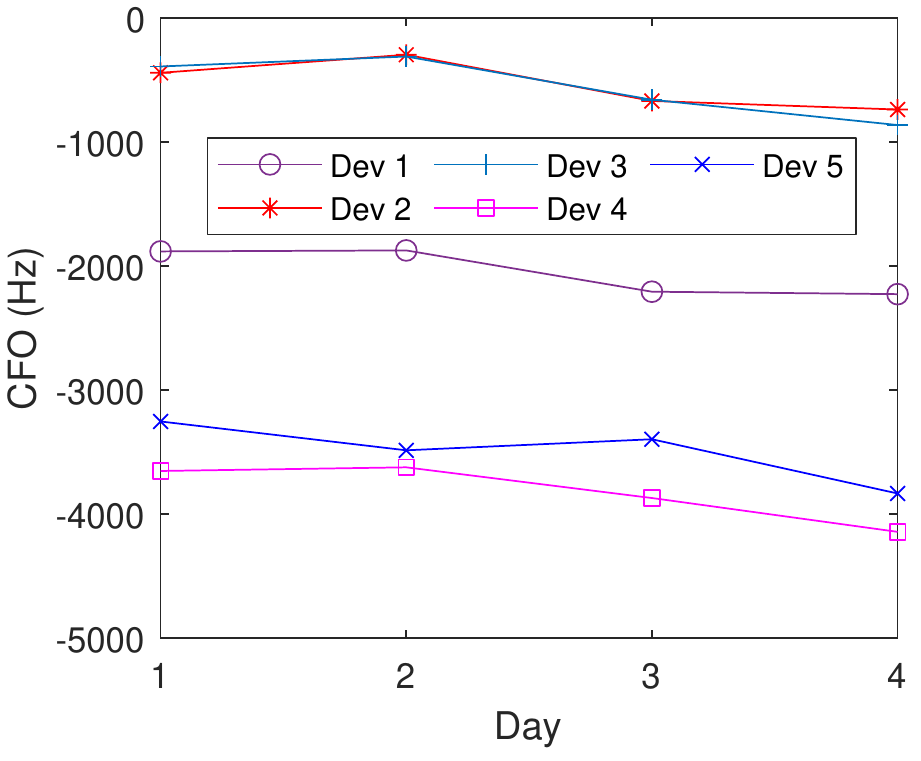}
		\label{fig:cfolongTime1_5}}
	\subfloat[]{\includegraphics[width=1.7in]{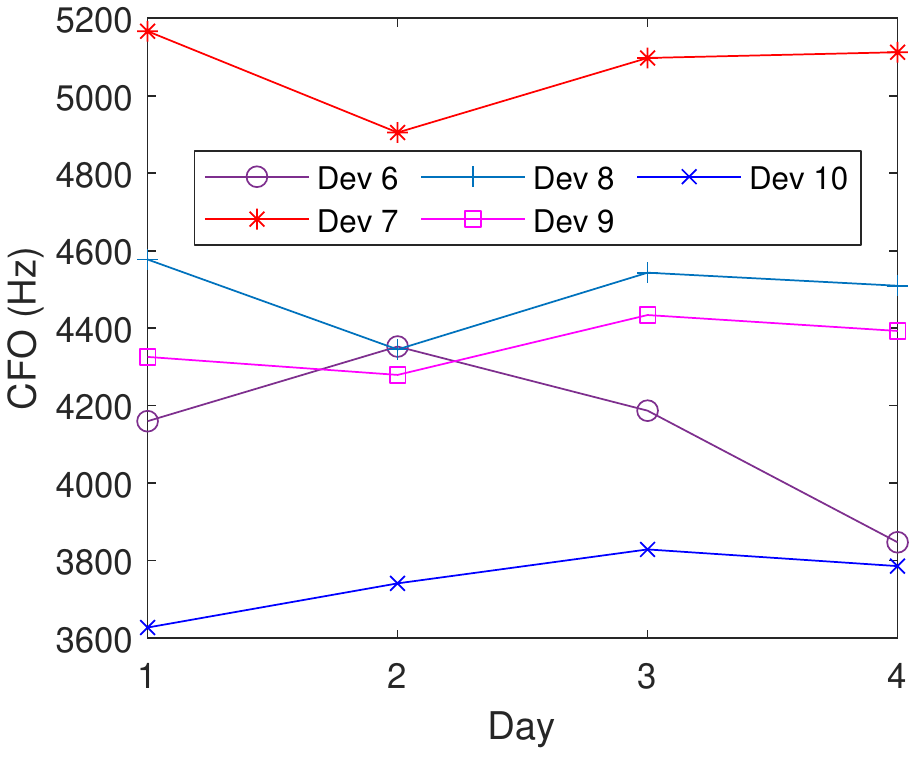}
		\label{fig:cfolongTime6_10}}
	\caption{Variations of average CFO within four days. (a) Dev 1-5. (b) Dev 6-10.}
	\label{fig:cfoLongTime}
\end{figure}

We further investigated CFO drifts over different days. 
We estimated the CFO of the packets collected on the same day and calculated the average value. The results are shown in Fig.~\ref{fig:cfoLongTime}. There is a non-negligible and unpredictable CFO change on different days. The drift is probably caused by the uncontrollable changes in room temperature. The long-time variation indicates that the classification accuracy may decrease when the training and test data are not collected on the same day since they probably have different CFOs. 

However, it can also be observed their average CFO remained relatively stable over the four days. While the time-varying CFO might not be suitable as a device identifier, it can be used to assist classification by ruling out devices whose estimated CFO deviates from the range too much.

\subsection{The Effect of CFO Drift on RFFI}\label{sec:expWired}

%In this subsection, a rigorous experiment was conducted to verify this conclusion: the performance of classifier will decrease significantly over time when without compensation, and CFO compensation can successfully solve this problem.
We carried out extensive experiments to evaluate CFO effects on the RFFI. We used the spectrograms of eight preambles and the CNN model shown in Fig.~\ref{fig:stftModel}.

The CNN was trained with the first 1,000 packets of each device (1,000$\times$10 packets in total) from Day 1 dataset, among which 90$\%$ were randomly selected for training and the rest 10$\%$ were for validation. Then we used another 1,000 packets of each device from Day 1 dataset to test the trained CNN classifier. For Day 2-4 datasets, the first 1,000 packets of each device were used as the test data. This allowed us to evaluate the trained CNN classifier with packets collected on four different days.

\begin{figure}[!t]
	%	\centering
	\subfloat[]{\includegraphics[width=1.7in]{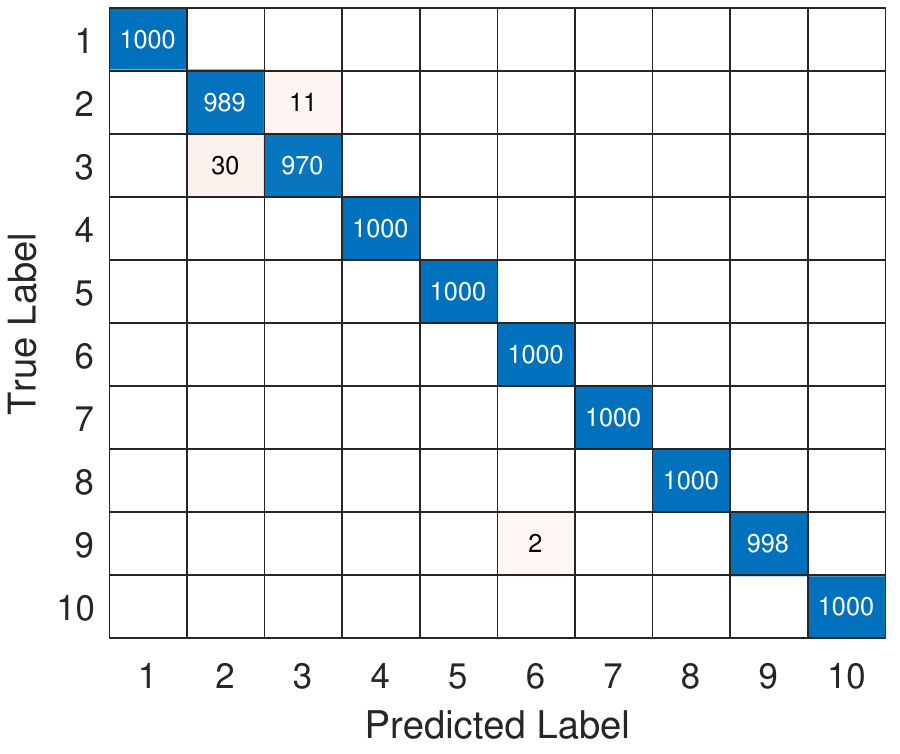}
		\label{fig:Day1Testuncompensated}}
	\subfloat[]{\includegraphics[width=1.7in]{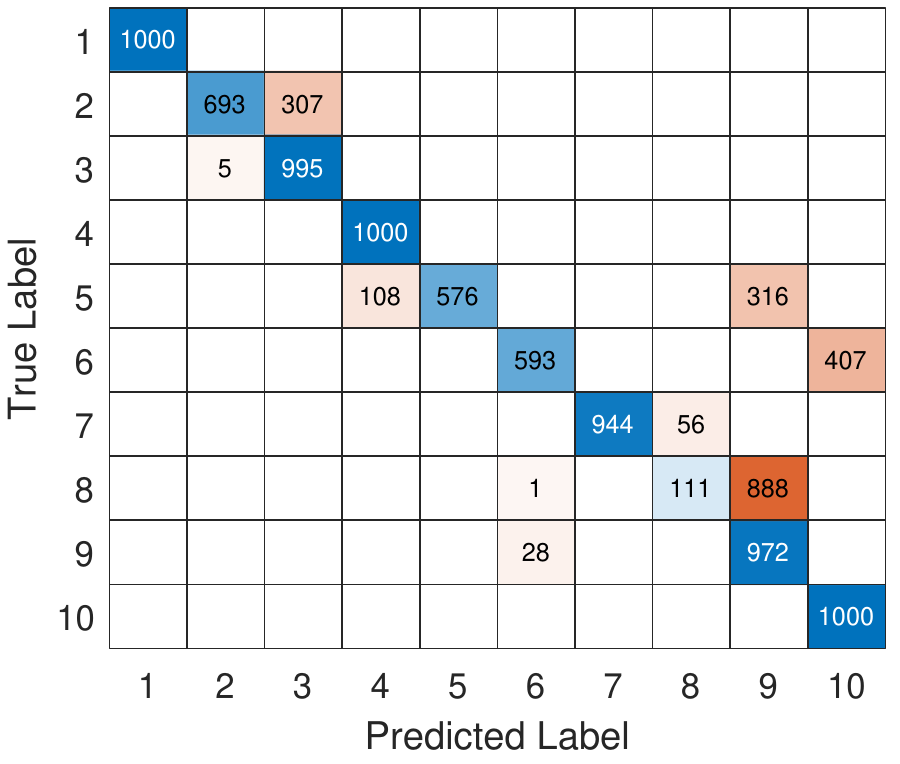}
		\label{fig:Day2Testuncompensated}}\\
	\subfloat[]{\includegraphics[width=1.7in]{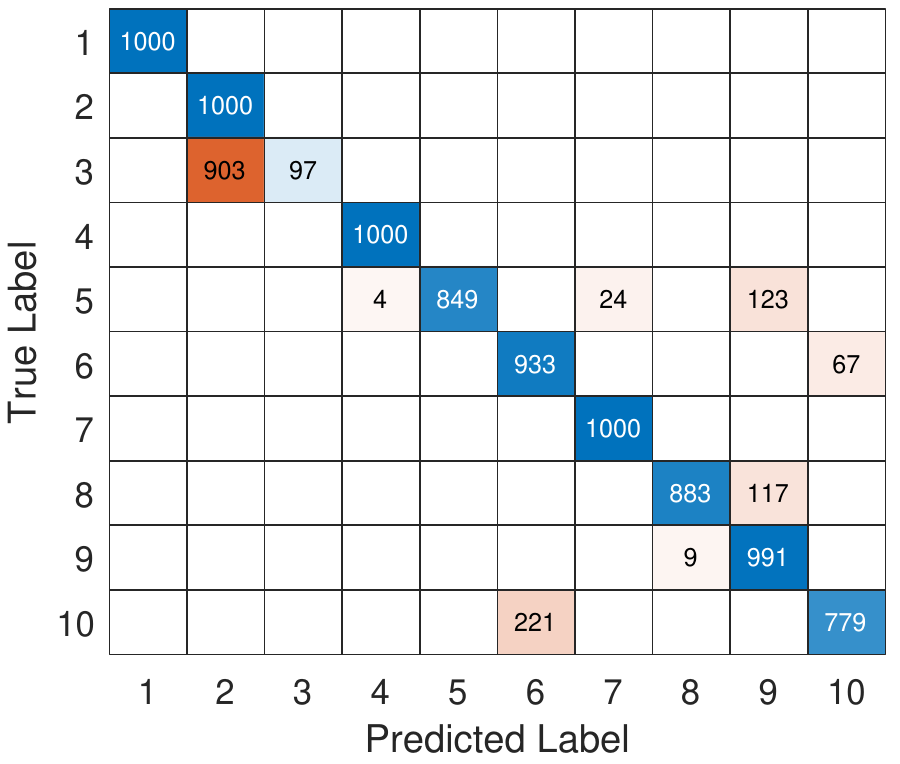}
		\label{fig:Day3Testuncompensated}}
	\subfloat[]{\includegraphics[width=1.7in]{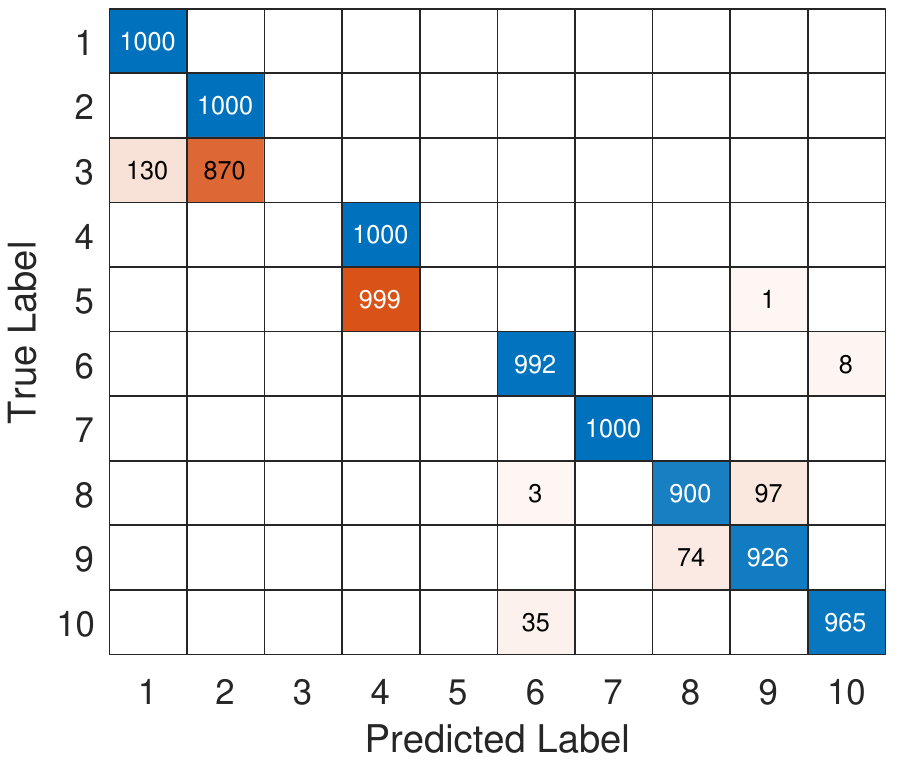}
		\label{fig:Day4Testuncompensated}}
	\caption{Experimental results without CFO compensation (CNN-only classifier). (a) Day 1 Training, Day 1 Test, overall accuracy: 99.57$\%$. (b) Day 1 Training, Day 2 Test, overall accuracy: 78.84$\%$. (c) Day 1 Training, Day 3 Test, overall accuracy: 85.32$\%$. (d) Day 1 Training, Day 4 Test, overall accuracy: 77.83$\%$.}
	\label{fig:expUncompensated}
\end{figure}

Fig.~\ref{fig:expUncompensated} shows the confusion matrices obtained by CNN-only classifier when CFO compensation was not applied. Figs.~\ref{fig:Day1Testuncompensated}, \ref{fig:Day2Testuncompensated}, \ref{fig:Day3Testuncompensated}, and \ref{fig:Day4Testuncompensated} represent the classification results when the test data was collected on Day~1, Day~2, Day 3, and Day 4, respectively. When the training and test sets were collected on the same day (Fig.~\ref{fig:Day1Testuncompensated}), the classification accuracy reached 99.57$\%$ which was almost no classification error. However, when the training and test data were collected on different days (Figs.~\ref{fig:Day2Testuncompensated}, \ref{fig:Day3Testuncompensated}, and \ref{fig:Day4Testuncompensated}), the classification results were unacceptable as several devices were completely misclassified, e.g., Dev 3 and Dev 5 in Fig.~\ref{fig:Day4Testuncompensated}. 

\begin{figure}[!t]
	\begin{center}
		\includegraphics[width = 3.4in]{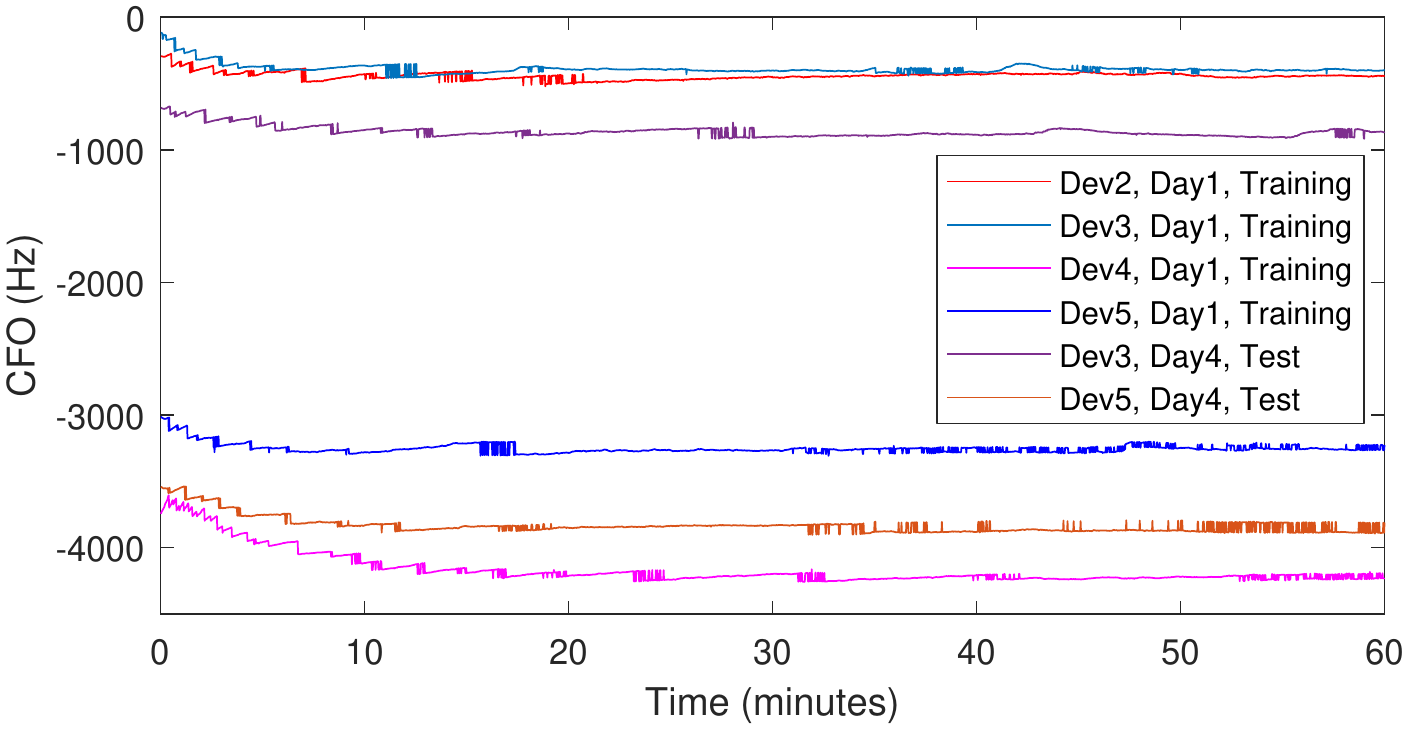}
		\caption{The comparison of CFO between the Day 1 training data and Day 4 test data. }
		\label{fig:CFOcompare}
	\end{center}
\end{figure}

The worst case happened on Day 4 (Fig.~\ref{fig:Day4Testuncompensated}) where the packets from Dev 3 and Dev 5 were completely misclassified. As shown in Fig.~\ref{fig:CFOcompare}, it can be observed that the CFO of Dev~3 and Dev 5 drifted by hundreds of hertz from Day 1 to Day~4. The CFO of Dev~3 in the test data (purple line) was closer to Dev 2 (red line) in the training data. Similarly, the CFO of Dev~5 (orange line) in the test data was closer to Dev 4 (pink line) in the training data. It is inferred that CFO drift was the main reason for performance degradation and a slight drift of CFO would cause the classifier to make a wrong decision.

\begin{figure}[!t]
	%	\centering
	\subfloat[]{\includegraphics[width=1.7in]{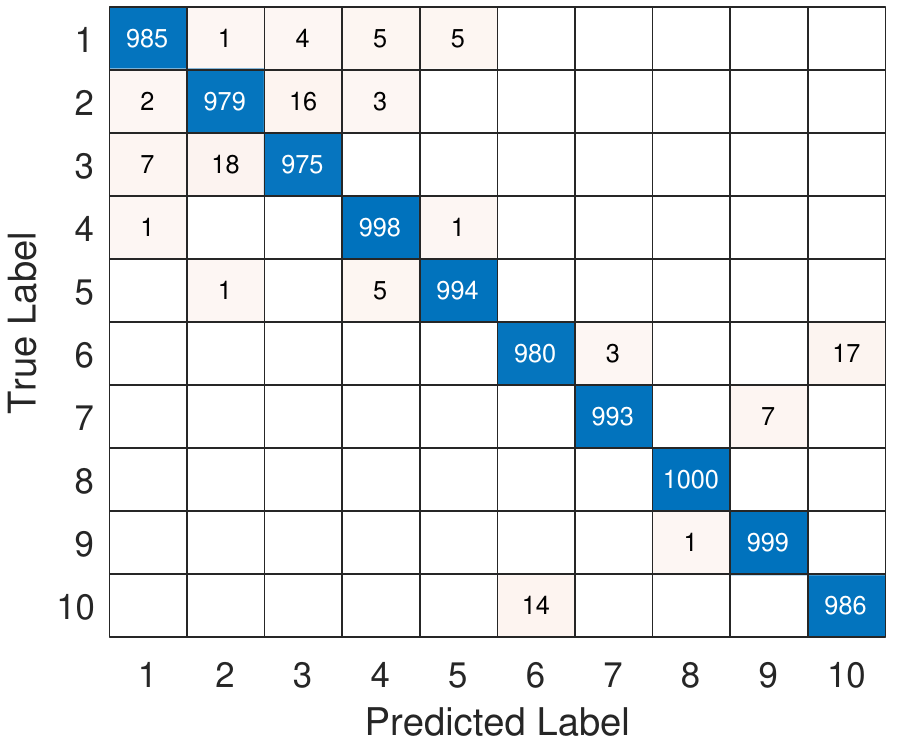}
		\label{fig:Day1Testcompensated}}
	\subfloat[]{\includegraphics[width=1.7in]{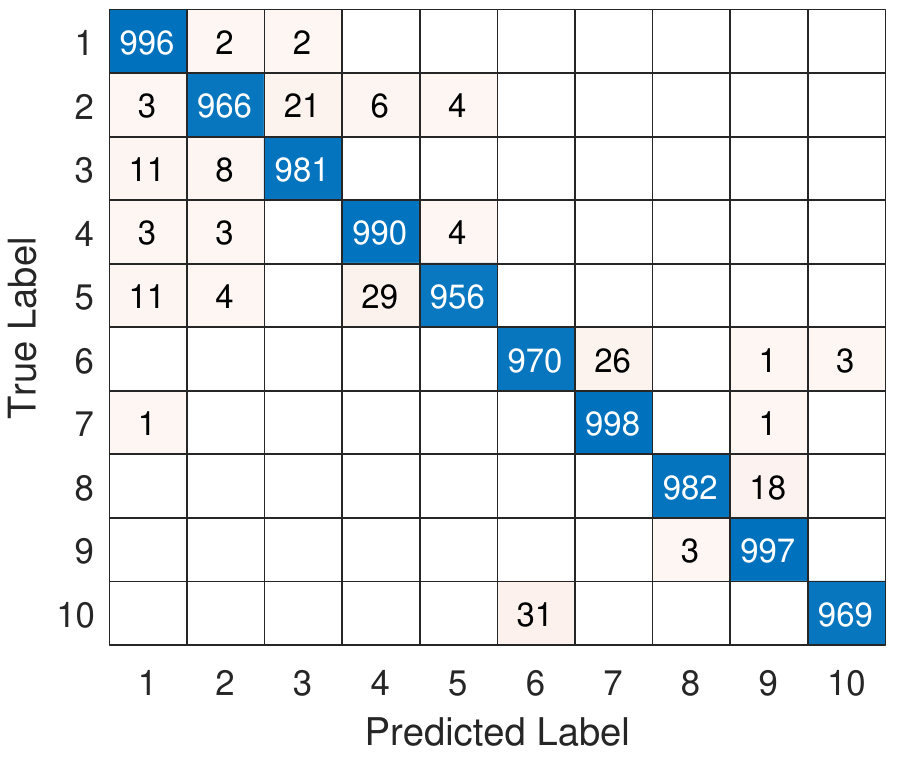}
		\label{fig:Day2Testcompensated}}\\
	\subfloat[]{\includegraphics[width=1.7in]{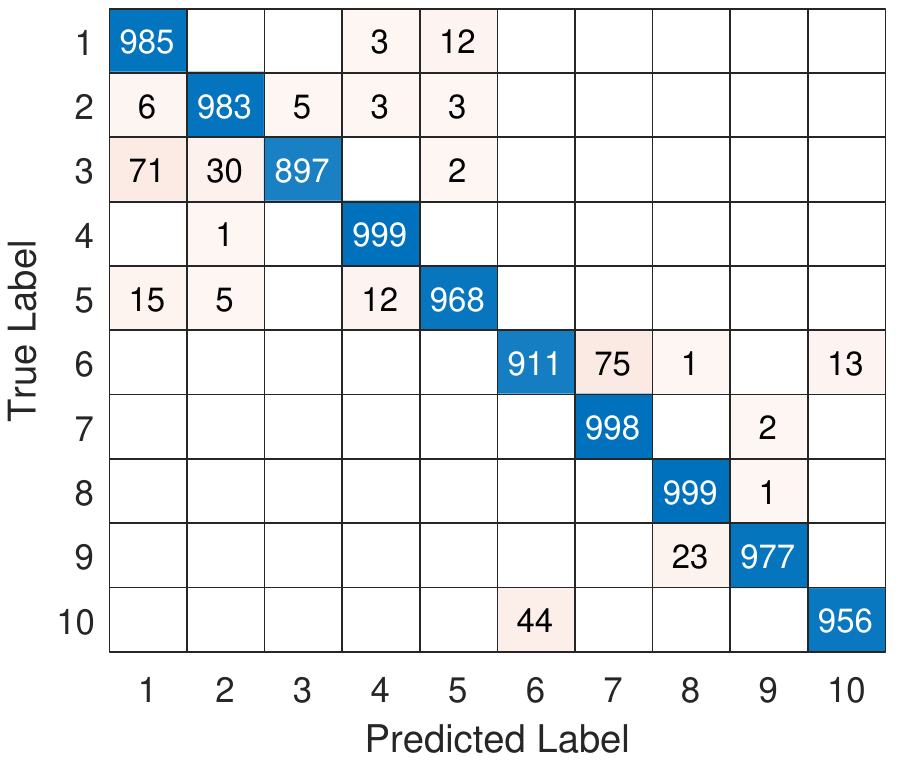}
		\label{fig:Day3Testcompensated}}
	\subfloat[]{\includegraphics[width=1.7in]{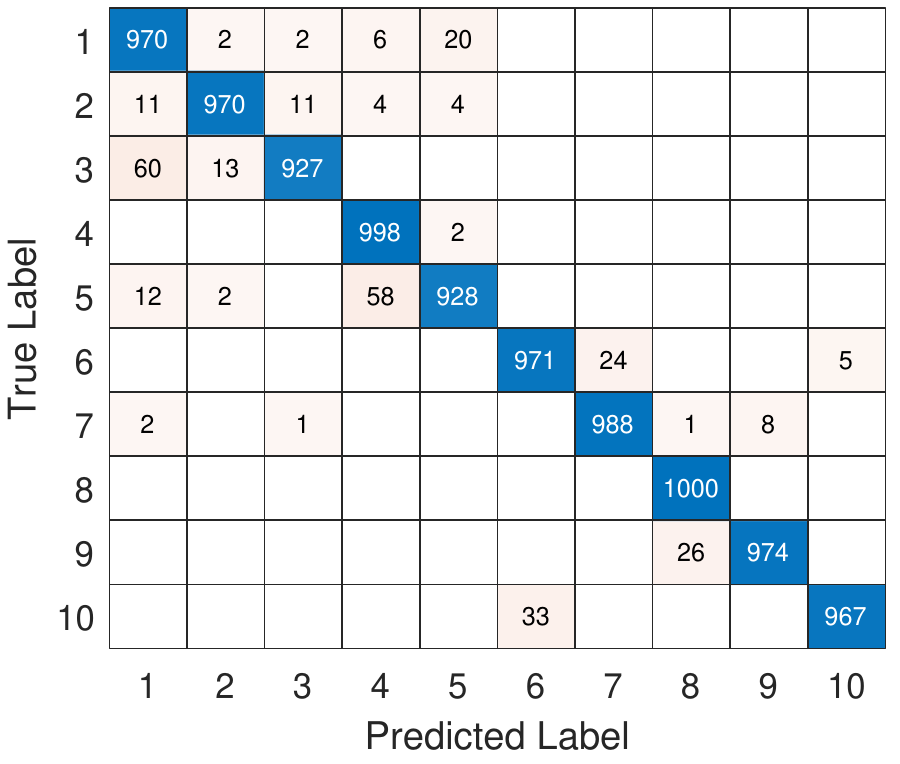}
		\label{fig:Day4Testcompensated}}
	\caption{Experimental results with CFO compensation (CNN-only classifier). (a) Day 1 Training, Day 1 Test, overall accuracy: 98.89$\%$. (b) Day 1 Training, Day 2 Test, overall accuracy: 98.05$\%$. (c) Day 1 Training, Day 3 Test, overall accuracy: 96.73$\%$. (d) Day 1 Training, Day 4 Test, overall accuracy: 96.93$\%$.}
	\label{fig:expCompensated}
\end{figure}

Fig.~\ref{fig:expCompensated} shows the confusion matrices obtained by CNN-only classifier after CFO compensation was applied. Compared with the results in Fig.~\ref{fig:expUncompensated}, there is no performance degradation after compensating the CFO, the accuracy always maintained above 96$\%$ on the four days. These results reveal that CNN can identify different devices with high accuracy after CFO compensation and performance degradation is significantly mitigated.
%removing the important CFO feature and the system does not suffer from the performance degradation problem any more after CFO compensation. 

\section{Experimental Evaluations in a Real Wireless Environment}\label{sec:expWireless}
In Section~\ref{sec:cfoDiscussion}, the LoRa DUT and USRP were connected using an attenuator, which allowed us to investigate the CFO effect on RFFI without the channel effect.
However, this is not a practical application scenario. Hence in this section, our proposed spectrogram-based RFFI system is evaluated in a real wireless environment. First, we compare the performance of three signal representations introduced in Section~\ref{sec:signal_representation}. Then we further evaluate the CFO effect in the wireless environment, and demonstrate that CFO compensation is an essential step in deep learning-based RFFI systems.
Finally, the calibration function of the proposed hybrid classifier is experimentally demonstrated. 

\subsection{Experimental Setup}
We increased the number of LoRa DUTs to 20 in this section. As shown in Table~\ref{tab:devInformation} and Fig.~\ref{fig:expDevices}, these LoRa devices were from four different manufacturers. The same USRP N210 platform was used as the receiver.
The LoRa DUTs and USRP platform were configured with the same parameters as described in Section~\ref{sec:setup}. The difference is that we shortened the transmission interval to 0.3 seconds to speed up signal collection.

The experiments were carried out in a typical indoor environment, with chairs and tables distributed in the room.
The distance between the LoRa DUT and USRP receiver was approximately three meters and there was line of sight (LOS) between them. 
We collected 2,000 packets continuously from each device for about 15 minutes.
%The signal collection of each device was repeated one by one. 
All the devices were placed at the same location and the environment was kept the same. Therefore, the same channel condition can be assumed for all the signal transmissions.

We used the first 1,000 packets of each device as training data, 90$\%$ of which were randomly selected for training and the rest 10$\%$ were for validation. 
The second 1,000 packets of each device were used as the test data to evaluate the RFFI system. The experimental results are presented in Table~\ref{tab:wirelessResults}. We analyzed the results from three aspects: the selection of signal representations, the impact of CFO in a wireless environment, and the calibration performance of our proposed hybrid classifier.

%\begin{table}[!t]
%  \centering
%  \caption{Experimental results}
%    \begin{tabular}{|l|r|r|r|r|}
%    \hline
%    \multirow{2}[4]{*}{Input type} & \multicolumn{2}{c|}{CNN-only} & \multicolumn{2}{c|}{Hybrid} \bigstrut\\
%\cline{2-5}          & \multicolumn{1}{L{1.3cm}|}{w/o CFO Comp.} & \multicolumn{1}{L{1.3cm}|}{\cellcolor[rgb]{ .816,  .808,  .808}w/ CFO Comp.} & \multicolumn{1}{L{1.3cm}|}{w/o CFO Comp.} & \multicolumn{1}{L{1.3cm}|}{\cellcolor[rgb]{ .816,  .808,  .808}w/ CFO Comp.} \bigstrut\\
%    \hline
%    IQ samples& 55.94\% & \cellcolor[rgb]{ .816,  .808,  .808}83.36\% & 58.28\% & \cellcolor[rgb]{ .816,  .808,  .808}92.01\% \bigstrut\\
%    \hline
%    FFT results & 49.18\% & \cellcolor[rgb]{ .816,  .808,  .808}87.36\% & 54.52\% & \cellcolor[rgb]{ .816,  .808,  .808}92.31\% \bigstrut\\
%    \hline
%    Spectrogram & 84.75\% & \cellcolor[rgb]{ .816,  .808,  .808}96.44\% & 85.73\% & \cellcolor[rgb]{ .816,  .808,  .808}97.61\% \bigstrut\\
%    \hline
%    \end{tabular}%
%  \label{tab:wirelessResults}%
%\end{table}%

\begin{table}[!t]
	\centering
	\caption{Experimental Results. Overall Classification Accuracy.}
	\begin{tabular}{|l|r|r|r|r|}
		\hline
		~ & \multicolumn{2}{c|}{CNN-only Classifier} & \multicolumn{2}{c|}{Hybrid Classifier} \bigstrut\\ \hline
		~          & \multicolumn{1}{L{1.3cm}|}{w/o CFO Comp.} & \multicolumn{1}{L{1.3cm}|}{\cellcolor[rgb]{ .816,  .808,  .808}w/ CFO Comp.} & \multicolumn{1}{L{1.3cm}|}{w/o CFO Comp.} & \multicolumn{1}{L{1.3cm}|}{\cellcolor[rgb]{ .816,  .808,  .808}w/ CFO Comp.} \bigstrut\\
		\hline
		IQ samples& 59.44\% & \cellcolor[rgb]{ .816,  .808,  .808}83.36\% & 59.45\% & \cellcolor[rgb]{ .816,  .808,  .808}92.01\% \bigstrut\\
		\hline
		FFT results & 51.62\% & \cellcolor[rgb]{ .816,  .808,  .808}87.36\% & 51.63\% & \cellcolor[rgb]{ .816,  .808,  .808}92.31\% \bigstrut\\
		\hline
		Spectrogram & 75.59\% & \cellcolor[rgb]{ .816,  .808,  .808}96.44\% & 75.59\% & \cellcolor[rgb]{ .816,  .808,  .808}97.61\% \bigstrut\\
		\hline
	\end{tabular}%
	\label{tab:wirelessResults}%
\end{table}%

\subsection{Selection of Signal Representations} \label{sec:comparisonWithBaseline}
We compare the classification accuracy of the three signal representations. The IQ/FFT-based CNN has a similar network structure with the spectrogram-based model hence a relatively fair comparison can be carried out.

As shown in Table~\ref{tab:wirelessResults}, when the CNN-only classifiers were used, the spectrogram-based model reached the highest accuracy of 96.44$\%$, while the  IQ and FFT-based model only reached 83.36$\%$ and 87.36$\%$, respectively. This shows that for LoRa signals whose frequency components are time-changing, device fingerprints can be detected more easily in the time-frequency domain.

Besides the classification results demonstrated in Table~\ref{tab:wirelessResults}, we found that the training time of our spectrogram-based model and the IQ/FFT-based model was about 20 minutes and one hour, respectively, when both were trained on the same PC. In addition to this, the loss of spectrogram-based model drops earlier and faster than the IQ/FFT-based model. This is another advantage of spectrogram-CNN model in terms of training costs.

%It is intuitive that there are many unique features hidden in the spectrograms. 

%\begin{itemize}
% 	\item [1)]
%    The spectrogram is more suitable as the input of CNN because CNN is originally designed for image classification, it is better at extracting the hidden features in images than time-series data.
% 	 
% 	\item [2)]
% 	Taking the spectrogram as network input can reveal more abundant time-frequency features of LoRa signals. LoRa devices exploit the linear chirps for communication, which are typical non-stationary signals. As shown in Fig.~\ref{fig:STFT}, the spectrogram obtained by STFT is an efficient approach to represent LoRa packets. Intuitively, there are many unique features hidden in the spectrograms. 
% 	
%\end{itemize}

%The results demonstrate that taking spectrograms as the input of CNN leads to the best classification accuracy, therefore STFT preprocessing is quite suitable for the RFF of LoRa devices.

%	\item [2)]

%\end{itemize}

\subsection{Impact of CFO drift}

As can be observed in Table~\ref{tab:wirelessResults}, when there was no CFO compensation, the accuracies of IQ, FFT, and spectrogram-based RFFI systems were only 59.44$\%$, 51.62$\%$ and 75.59$\%$, respectively. After CFO compensation was applied, the corresponding accuracies significantly increased to 83.36$\%$, 87.36$\%$ and 96.44$\%$, respectively. 

\begin{figure}[!t]
	\begin{center}
		\includegraphics[width = 3.4in]{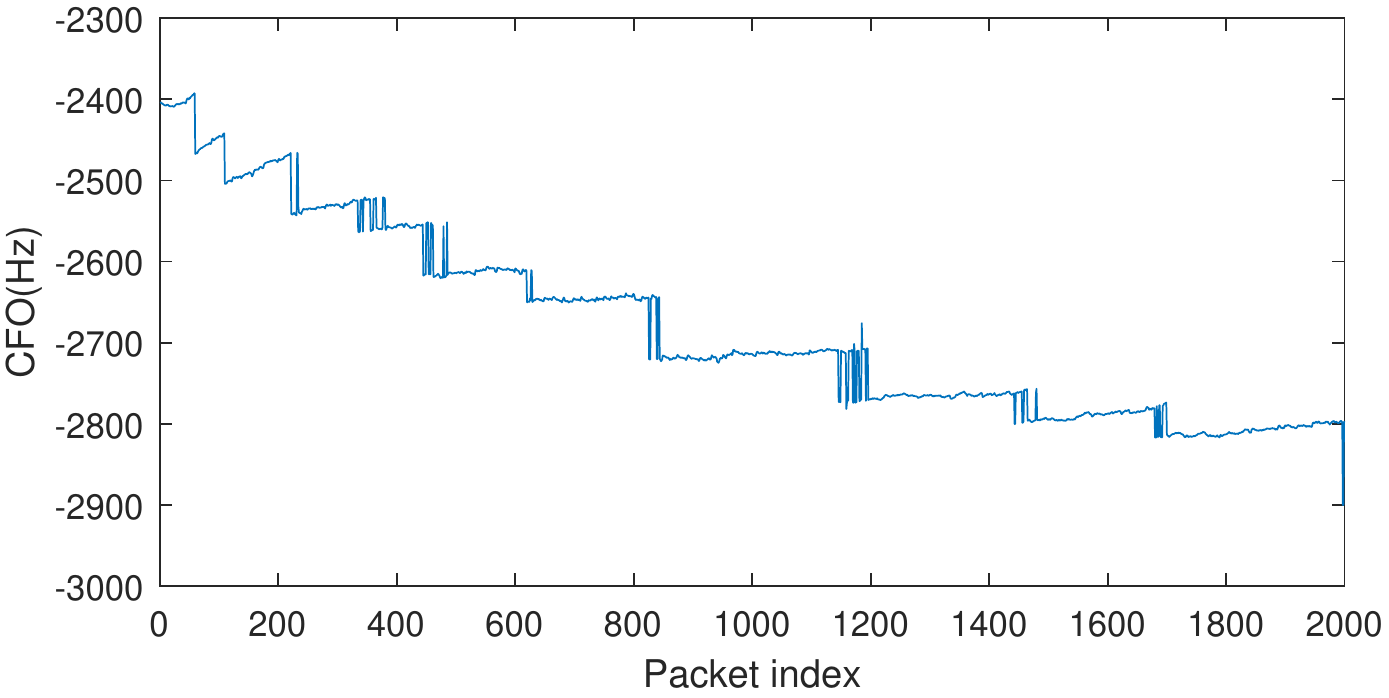}
		\caption{CFO of each packet in the dataset of Dev 1. }
		\label{fig:Dev1CFO}
	\end{center}
\end{figure}
%Take the dataset of Dev 1 as an example to explain the terrible results without CFO compensation. Fig.~\ref{fig:Dev1CFO} shows the CFO of each packet collected from Dev 1, it is clear that the CFO decreased continuously after the device was powered on, probably due to the temperature variations caused by self-heating. In the experiments, we used packet 1-1,000 to train the CNN and packet 1,001-2,000 to evaluate its performance. However, the packets in the test set have different CFO with those in the training set, in other words, the test data has different underlying distributions with the training data.  In machine learning tasks, the training set and test set are often required to have the same, at least similar data distributions, otherwise, the trained model will face severe generalization problem. Therefore, we infer that the poor performance without CFO compensation results from the different distributions of the training set and the test set, i.e, CFO drift, and CFO compensation can well mitigate this problem.

We take the dataset of Dev 1 as an example to explain the results without CFO compensation. Fig.~\ref{fig:Dev1CFO} shows the CFO of each packet collected from Dev 1, and presents a similar pattern with Fig.~\ref{fig:cfoShortTime} that the CFO decreased continuously after the device was powered on. In the wireless experiments, we used packet 1-1,000 to train the CNN and packet 1,001-2,000 to evaluate its performance. However, the packets in the test set have different CFOs with those in the training set. In other words, the test data has different distributions from the training data. 
In machine learning tasks, the training and test sets are often required to have the same, at least similar data distributions, otherwise, the trained model will face severe generalization problems. This might also be the reason for the low classification accuracy in~\cite{robyns2017physical} when CFO compensation is not involved.
\begin{figure}[!t]
	\centering
	\subfloat[]{\includegraphics[width=1.7in]{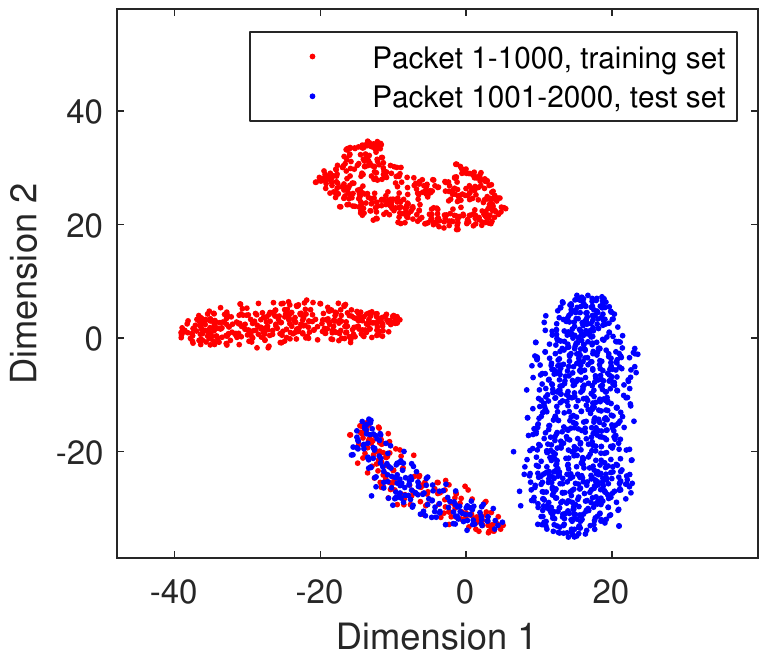}
		\label{fig:tSNE_uncompensated_wireless}}
	\subfloat[]{\includegraphics[width=1.7in]{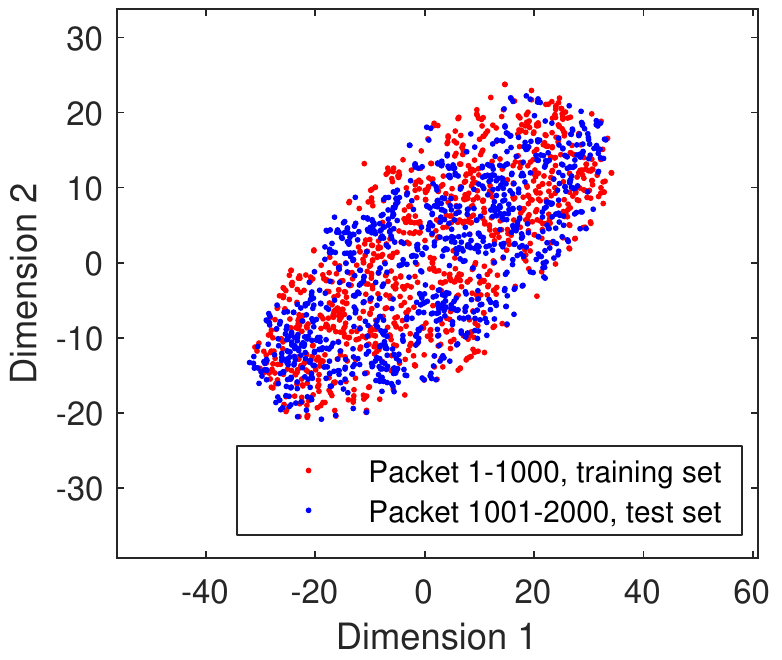}
		\label{fig:tSNE_compensated_wireless}}
	\caption{t-SNE visualization of the training and test sets of Dev1. (a) Without CFO compensation. (b) With CFO compensation.}
	\label{fig:tSNE_wireless}
\end{figure}

Then we verify the argument that the training data and test data have different distributions using the well-known t-SNE visualization algorithm. The visualization result is shown in Fig.~\ref{fig:tSNE_wireless}, in which each point represents a packet collected from Dev 1. There are 2,000 points in total and the red points represent packets 1-1,000 (training data) and blue points represent packets 1,001-2,000 (test data). From Fig.~\ref{fig:tSNE_uncompensated_wireless} it can be observed that there are four distinct clusters when there is no CFO compensation, which indicates that the training data and test data have different features/distributions. In contrast, as shown in Fig.~\ref{fig:tSNE_compensated_wireless}, the blue and red points are mixed after CFO compensation and cannot be separated intuitively. This is what we expected because the features of each device should be time-invariant, i.e., the first 1,000 packets should have the same features with the second 1,000 packets, which leads to overlapping in the visualization.

%
%The results strongly demonstrate that CFO compensation is indispensable for deep learning-based RFFI system, regardless of the input type. 

%\begin{figure}[!t]
%	\centering
%	\subfloat[]{\includegraphics[width=3.4in]{Pictures/Day2_uncompensated.pdf}
%		\label{fig:wirelessUncompensated}}\\
%	\subfloat[]{\includegraphics[width=3.4in]{Pictures/Day2_compensated_hybrid_2.pdf}
%		\label{fig:wirelessCompensated}}
%	\caption{Classification results of the spectrogram-based model when the training and test sets were collected on different days. (a) Without CFO compensation, overall accuracy: 77.48$\%$. (b) With CFO compensation, overall accuracy: 96.35$\%$.}
%	\label{fig:systemStability}
%\end{figure}

\subsection{Effectiveness of The Hybrid Classifier}
The hybrid classifier introduced in Section~\ref{sec:hybrid} calibrates the softmax output of CNN according to the estimated CFO. As shown in Fig.~\ref{fig:cfoLongTime}, the CFO varies over different days and some devices may have similar CFOs, hence it cannot be used as a fingerprint to identify numerous low-cost IoT devices. However, the average values of CFO stay relatively stable in a small range, which can be used for calibration to rule out predictions whose estimated CFO is much different from the reference one.

As shown in Table~\ref{tab:wirelessResults}, it can be observed that the hybrid classifier can increase the accuracy for all the three signal representations. The most significant improvement was the input type of IQ samples after applying CFO compensation, the accuracy with hybrid classifier for IQ data reached 92.01$\%$ while the accuracy using CNN-only classifier was 83.36$\%$, which was an 8.65$\%$ improvement. For the signal representation of FFT results, there was an accuracy improvement from 87.36$\%$ to 92.31$\%$, and for the spectrogram, the accuracy increased from 96.44$\%$ to 97.61$\%$. 

It is also observed that the hybrid classifier does not work when there was no CFO compensation. This is reasonable because the CFO has contributed to the prediction when compensation is not involved and the hybrid classifier cannot provide additional helpful information.

\section{Conclusion}\label{sec:conclusion}
In this paper, we proposed a spectrogram-based RFFI system and carried out extensive experimental evaluations. We used 20 LoRa devices of four models as the DUTs and a USRP N210 SDR as the receiver. 
Firstly, as LoRa uses chirp modulation, we employed spectrogram to represent the time-frequency characteristics of LoRa signals. We found that using spectrogram can achieve a better classification accuracy compared to the  IQ samples and FFT results.
Secondly, we experimentally found that CFO is not stable as it was varying over time probably due to temperature changes. Hence it will compromise the system stability. 
CFO compensation was experimentally found to be effective in mitigating the performance degradation. 
Finally, we proposed a hybrid classifier that calibrates the softmax output of CNN using the estimated CFO. Although CFO is varying over time, its average value stays relatively stable over days. CFO must be compensated to avoid performance degradation but is helpful to rule out predictions when the estimated CFO deviates greatly from the reference CFO.
%Our results show that the proposed hybrid classifier brought 1.51\% and 8.65\% improvement to the classification accuracy when the spectrogram and the raw IQ data were used as network inputs, respectively. 
Our proposed RFFI system finally achieved a classification accuracy of 97.61\% in distinguishing 20 LoRa devices in real wireless environments. 

%Our future work will focus on eliminating channel effects on the RFFI system.
	
\section*{Acknowledgement}
The work was in part supported by the UK Royal Society Research Grants under grant ID RGS\slash R1\slash 191241 and national key research and development program of China under grant ID 2020YFE0200600.

\bibliographystyle{IEEEtran}
\bibliography{IEEEabrv,mybibfile}

%	Biography text here.
%\end{IEEEbiography}
%	Biography text here.
%	Biography text here.
	
\end{document}